\documentclass{aa} 
\usepackage{natbib}
\usepackage{graphicx}
\usepackage{ulem}
\usepackage[varg]{txfonts}
\usepackage{natbib}
\usepackage[colorlinks=true,citecolor=blue,linkcolor=blue]{hyperref}

\graphicspath{{./figs/}}

\usepackage[dvipsnames]{xcolor}

\newcommand{\cyg}{\mbox{Cygnus~X-1}\xspace}

\begin{document}

\title{The spectral-timing analysis of \cyg with Insight-HXMT}
\titlerunning{The spectral-timing analysis of \cyg with Insight-HXMT}

\author{
M.~Zhou\inst{1}\and
V.~Grinberg\inst{2}\and
Q.-C.~Bu\inst{1}\and
A.~Santangelo\inst{1}\and
F.~Cangemi\inst{3}\and
C.~M.~Diez\inst{1}\and
O.~K\"onig\inst{4}\and
L.~Ji\inst{5}\and
M.~A.~Nowak\inst{6}\and
K.~Pottschmidt\inst{7, 8}\and
J.~Rodriguez\inst{9}\and
J.~Wilms\inst{4}\and
S.~Zhang\inst{10, 11}\and
J.-L.~Qu\inst{10, 11}\and
S.-N.~Zhang\inst{10, 11}
}

\institute{Institut f\"ur Astronomie und Astrophysik, Universität T\"ubingen, Sand 1, 72076 T\"ubingen, Germany \\
  \texttt{menglei.zhou@astro.uni-tuebingen.de}
\and European Space Agency (ESA), European Space Research and Technology Centre (ESTEC), Keplerlaan 1, 2201 AZ Noordwijk, The Netherlands 
\and Laboratoire de Physique Nucl\'eaire et des Hautes \'Energies, Sorbonne Universit\'e, 4 Place Jussieu, 75005 Paris, France 
\and Dr. Karl Remeis-Sternwarte and ECAP, Friedrich-Alexander-Universit\"at Erlangen-N\"urnberg, Sternwartstr. 7, 96049 Bamberg, Germany 
\and School of Physics and Astronomy, Sun Yat-Sen University, 519082 Zhuhai, China 
\and Physics Department, Washington University CB 1105, St Louis, MO 63130, USA
\and CRESST and Astroparticle Physics Laboratory, NASA Goddard Space Flight Center, Greenbelt, MD 20771, USA 
\and Department of Physics and Center for Space Sciences and Technology, University of Maryland, Baltimore County, Baltimore, MD 21250, USA 
\and Universit\'e Paris-Saclay, Universit\'e Paris Cit\'e, CEA, CNRS, AIM de Paris-Saclay, 91191 Gif sur Yvette, France 
\and Key Laboratory for Particle Astrophysics, Institute of High Energy Physics, Chinese Academy of Sciences, 100049 Beijing, China
\and University of Chinese Academy of Sciences, Chinese Academy of Sciences, 100049 Beijing, China
}

\date{ -- / --}

\abstract{
\cyg, as the first discovered black hole binary, is a key source for understanding the mechanisms of state transitions, and the scenarios of accretion in extreme gravity fields. We present a spectral-timing analysis of observations taken with the Insight-HXMT mission, focusing on the spectral-state dependent timing properties in the broad energy range of 1--150\,keV, thus extending previous RXTE-based studies to both lower and higher energies. Our main results are the following: a) We successfully use a simple empirical model to fit all spectra, confirming that the reflection component is stronger in the soft state than in the hard state; b) The evolution of the total fractional root mean square (rms) depends on the selected energy band and the spectral shape, which is a direct result of the evolution of the power spectral densities (PSDs); c) In the hard/intermediate state, we see clear short-term variability features and a positive correlation between central frequencies of the variability components and the soft photon index $\Gamma_1$, also at energies above 15\,keV. The power spectrum is dominated by red noise in the soft state instead. These behaviors can be traced to at least 90\,keV; d) The coherence and the phase-lag spectra show different behaviors dependent on different spectral shapes. 
}


    \maketitle

\section{Introduction}

Black hole binaries (BHBs), are systems consisting of a black hole accreting material from a donor star. \citep[For a review on BHBs, see e.g.,][]{remillard_2006, done_2007, dunn_2010} The radiation produced by the accreting process is bright in the X-rays, exhibiting a rich phenomenology in both spectral and timing domains. Most BHBs are transient, and exhibit luminous outburts. \cyg is one of the few persistent systems~\citep{bowyer_1965, bolton_1972, webster_1972}. \citet{tananbaum_1972} first reported a global change in the X-ray spectra of \cyg, with a simultaneous switching-on of the jet emission. It was later realized that this corresponded to a transition from the soft to the hard state. 

The X-ray spectrum of the hard state is characterized by a (cut-off) powerlaw component with a photon index $\Gamma \lesssim 2.0$ and with weak or non-detectable thermal black body emission from the accretion disk~\citep[e.g.,][]{tomsick_2008}. The powerlaw emission is presumably produced via Comptonization by a hot ($k T_{\mathrm{e}} \approx 100$\,keV) plasma~\citep{haardt_1993, dove_1997, zdziarski_2003, ibragimov_2005}. In contrast, the spectrum in the soft state is dominated by thermal emission with a characteristic temperature of several hundred eV and a steep powerlaw with $\Gamma \gtrsim 2.7$. In between these two states, the intermediate state is defined, which exhibits spectral characteristics intermediate to the hard and the soft state, but with distinct timing features. 
BHB transients spend most of their time in the quiescent state, which is extraordinarily faint and hard~\citep[$1.5 \lesssim \Gamma \lesssim 2.1$, see e.g.,][]{plotkin_2013}.
However, they usually trace the so-called \texttt{q}-track in the Hardness-Intensity Diagram (HID) during the outburst, moving from the  quiescent state, to the hard state, the intermediate state, and the soft state ~\citep{homan_2001, fender_2006}.


 Timing properties of BHBs also show distinct behaviors in different states~\citep{wijnands_1999}. In the hard and intermediate state, we see a strong intrinsic variability with the central frequencies mostly located in the 0.1--10\,Hz range, often referred to as ``low frequency quasi-periodic oscillations'' (LFQPOs) if it is a narrow and prominent feauture~(see e.g.,~\citealt{done_2007}, and for a recent review see~\citealt{ingram_2019}). The variability is high ($\sim 20$\% rms) in the hard state~\citep{belloni_2010}. The central frequencies of the variability components have been found to have a strong correlation with the spectral state~\citep[see][and the references therein]{remillard_2006}. However, in the soft state, the power spectrum is dominated by the red noise with slope $f^{-1}$ below 10\,Hz and a cut-off tail at higher frequencies~\citep[e.g.,][]{gilfanov_2000, axelsson_2005}. During the hard-to-intermediate transition, we often observe a continued growth of the averaged hard lags (i.e., the hard photons lag the soft photons) between two coherent energy bands~\citep{cui_1997_temporal, pottschmidt_2000, reig_2018}. The time-lags reach the maximum just before completing the transition to the soft state, and then drop to $\sim$ 0 seconds in the soft state~\citep{boeck_2011, grinberg_2014, altamirano_2015}. 

The physics of the state transition is not fully understood yet. A widely-used scenario to explain the state transition behaviors is the ``truncated disk/hot inner flow model''~\citep{ichimaru_1977, narayan_1995, narayan_2008}. The outer part of the disk is described by a standard optically-thick accretion disk~\citep{shakura_sunyaev_1973}, where the emission produced by the disk is modeled by the thermal black body radiation. The disk is truncated at some radius. The inner part is substituted by an optically-thin hot flow, which corresponds to the ``corona'' that produces the non-thermal powerlaw emission~\citep[for a detailed review, see e.g.,][]{done_2007}. The nature of the corona is an open question as well, with some theories associating the corona with the base of a jet~\citep{markoff_2005}. State transitions from hard to soft happen when the inner edge of the standard disk moves inward and the size of the optically-thin hot flow shrinks. 
This model can partly explain the phenomena we observe during the state transition, but the observations support conflicting scenarios; on the one hand, e.g., \citet{gierlinski_2004}, \citet{penna_2010}, \citet{steiner_2010}, \citet{zhu_2012} and \citet{zdziarski_2022} support this scenario by obtaining large truncated radii in the hard state, with the truncated radii extending to the innermost stable circular orbits (ISCOs) when the transitions to the soft state are completed; on the other hand, this model is contested by other measurements of the inner edge of the disk, either by fitting the broadened iron K$\alpha$ line~\citep[see e.g.,][]{garcia_2015, garcia_2019, sridhar_2020} or even the disk component itself~\citep{miller_2006, rykoff_2007, reis_2009, reis_2010, reynolds_2010}. 


Spectral-timing analysis might help to better understand the physics (and geometry) of state transitions. For instance, \citet{axelsson_2005} showed the evolution of the power spectra in different states of \cyg and the correlation between two central frequencies of the variability component; \citet{gierlinski_2005} showed the correlation between the total fractional root mean square (rms) and the photon energy spectra; \citet{cassatella_2012} simultaneously modeled the energy spectra and the frequency-dependent time-lags; \citet{arevalo_2006} instead tried using a numerical implementation to reproduce the spectral-timing behaviors seen in BHBs. 


In this paper, we focus on \cyg, the first confirmed black hole candidate, with a supergiant O9.7 Iab donor star (HDE~226868). Using VLBA observations, \citet{science_2021} recently estimated a black hole mass of $M_1 = (21.2 \pm 2.2) \, M_{\odot}$, a distance $D = 2.22^{+0.18}_{-0.17}$\,kpc, and the mass of the donor star to be $M_2 = 40.6^{+7.7}_{-7.1} \, M_{\odot}$. The compact object in \cyg is a fast-rotating black hole, with a dimensionless spin parameter $a^{*} > 0.97$ (\citealt{fabian_2012, parker_2015, zhao_2021}, but see also e.g., \citealt{walton_2016}). Previous spectral-timing analyses on \cyg based on the \textsl{RXTE}/PCA data, e.g., \citet{pottschmidt_2003}, \citet{shaposhnikov_2006}, \citet{axelsson_2006}, and \citet{klein-wolt_2008}, studied the energy-independent evolution of the timing properties with different spectral shapes. \citet{boeck_2011} and \citet{grinberg_2014} studied the energy-dependent timing behavior based on abundant X-ray data that covered all the states of \cyg, in the energy range up to $\sim 15$\,keV. In this work we extend previous studies, using wide-band data provided by the HXMT satellite, and exploring the spectral and temporal behaviors of \cyg to a broader energy range, in particular covering energies above 15\,keV where the contribution from reflection becomes important. 


The remainder of this paper is structured as follows: in Sect.~\ref{sect:obs}, we present the long-term behaviors of the source and an overview of the HXMT data we analyze in this paper. We show the results of the spectral analysis in Sect.~\ref{sect:spec}, and the timing analysis extending to 1--150\,keV range in Sect.~\ref{sect:timing}. We discuss the obtained results of our timing analysis in Sect.~\ref{sect:discussion_1}, and make an extensive probe of the accreting scenario in Sect.~\ref{sect:discussion_2}, which is important for understanding BHB systems but remains hitherto controversial. The summary and outlook are in Sect.~\ref{sect:sum}. 

\section{Observations and Data}\label{sect:obs}

The Insight--Hard X-ray Modulation Telescope, also short for Insight--HXMT or HXMT, is China's first X-ray astronomical satellite launched on June 15, 2017~\citep{zhang_2014, zhang_2020}. It has a broad energy band ranging nominally from 1 to around 250\,keV, realized by three collimated telescopes: the Low Energy (LE) telescope, whose energy range covers the 1--15\,keV band~\citep{chen_2020_low}; the Medium Energy (ME) telescope, covering 5--30\,keV~\citep{cao_2020_medium}; and the High Energy (HE) telescope, covering the 25--250\,keV band~\citep{liu_2020_high}. In particular, HXMT's fast temporal response and large effective area, even at higher energies, provide us with the possibility of detailed spectral-timing studies above 15\,keV. 

\begin{figure*}
    \centering
    \includegraphics[width = 0.99\textwidth]{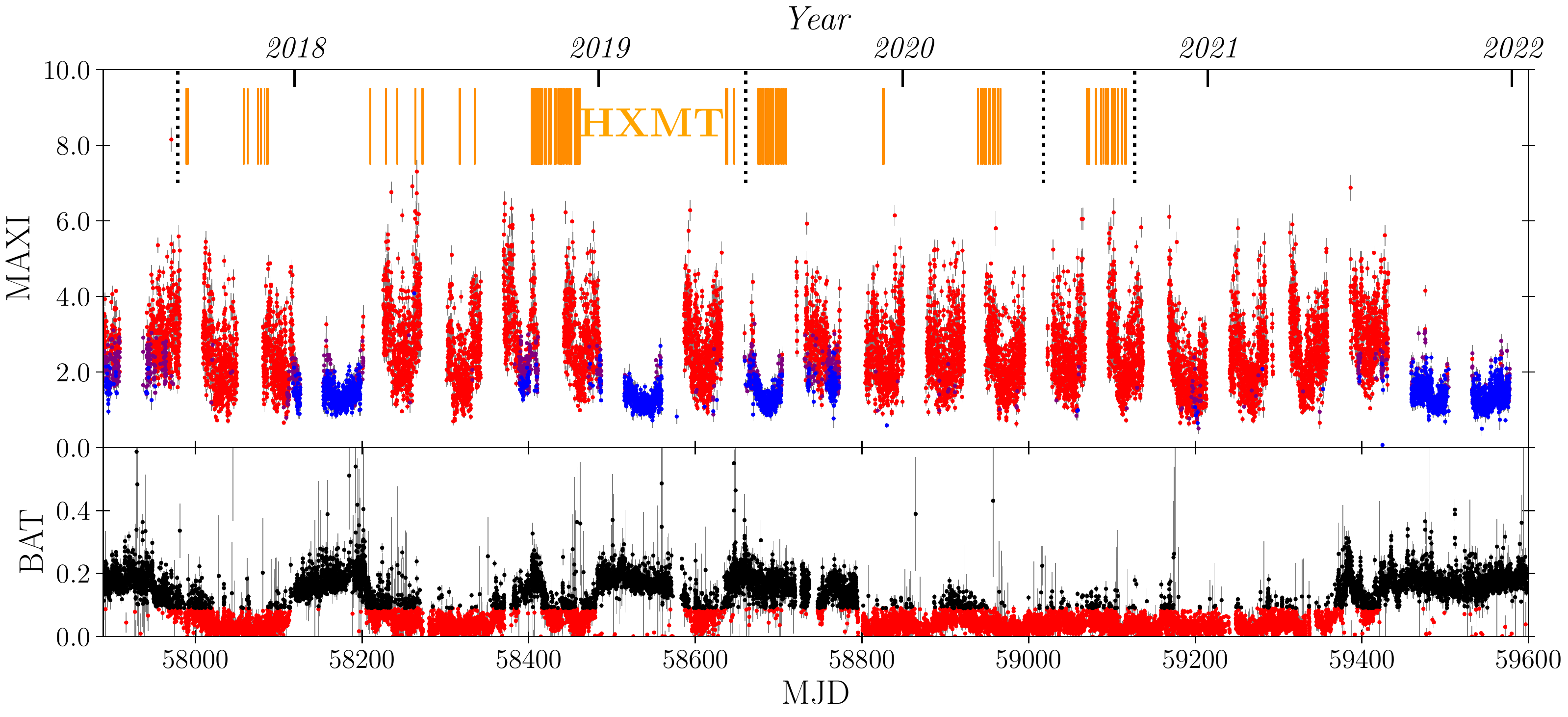}
    \caption{The long-term behavior of \cyg measured by all-sky monitors MAXI and \textit{Swift}-BAT. We follow the same approaches of state definition proposed by \citet{grinberg_2013}. In the \textit{upper panel}, the hard/intermediate/soft states are denoted by blue/purple/red points, respectively. In the \textit{lower panel}, the unclassified/soft states are plotted with black/red points. The duration of HXMT observations on \cyg is shown in dark orange stripes on the top. }
    \label{f-lc}
\end{figure*}

We use three series of observations. The proposal IDs and the dates of observations are listed in Table~\ref{tab:proposal-id}. The duration of these observations is shown in Fig.~\ref{f-lc}, along with the light curves of \cyg provided by two current all-sky monitors, MAXI~\citep{maxi} and \textit{Swift}-BAT~\citep{bat}. Different states of the source are indicated by different colors determined according to \citet{grinberg_2013}. We observe that \cyg has been mostly in the soft state since 2018, but often moves between the hard and soft state. 

\renewcommand{\arraystretch}{1.2}
\begin{table}
    \caption{The start dates and end dates of the HXMT observations. }\label{tab:proposal-id}
    \centering
    \begin{tabular}{lll}
        \hline\hline
        Proposal ID & Start Date~\tablefootmark{a} & End Date~\tablefootmark{a} \\
        \hline
        P0101315 & 2017-08-24 (57989) & 2019-06-12 (58646) \\
        P0201012 & 2019-07-11 (58675) & 2020-04-27 (58966) \\
        P0305079 & 2020-08-08 (59069) & 2020-09-25 (59117) \\
        \hline
    \end{tabular}
    \tablefoot{
    \tablefoottext{a}{The date format is yyyy-mm-dd (MJD).}}
\end{table}
\renewcommand{\arraystretch}{1.0}

\subsection{Data extraction}

We use the pipeline based on Insight--HXMT Data Analysis Software (\texttt{HXMTDAS}) v2.04 along with the latest calibration database v2.05 to process the observational data. The Good Time Intervals (GTIs) are selected according to the following criteria: the elevation angle (ELV) larger than $10^\circ$, the geometric cut-off rigidity (COR) larger than 8\,GeV, the offset angle from the pointing direction (ANG\_DIST) smaller than $0.04^\circ$, and at least 300 seconds before and after the South Atlantic Anomaly (SAA) passage. For the LE telescope, we adopt an additional criterion that the elevation angle for the bright Earth is larger than $30^\circ$. The individual detectors of the HXMT instruments have different field of views (FoVs). Given that \cyg is a bright source, we use photon events generated by the detectors with a small FoV for the following analysis. 

In our analysis, we consider those GTIs that are covered by all three instruments and that are longer than the individual segment used for timing analysis, i.e. longer than 16\,s. In our timing analyses, we thus used a total 145 exposures, for which we have obtained energy spectra. We use \texttt{ISIS 1.6.2}~\citep{houck_2000}, which allows access to the defined models in \texttt{Xspec}~\citep{arnaud_1996}, to perform the spectral fitting. The energy bands for the spectral analyses are 2--10\,keV for LE, 10--30\,keV for ME, and 28--120\,keV for HE. We use the tools \texttt{LEBKGMAP}, \texttt{MEBKGMAP}, and \texttt{HEBKGMAP} provided by the HXMT team to 
estimate the instrumental background~\citep{liao_2020_backgrounda, liao_2020_backgroundb, guo_2020_background}. 

The light curves for timing analyses are generated by the corresponding event lists from the detectors with a small FoV directly, processed with \texttt{Stingray}~\citep{bachetti_2021, huppenkothen_2019_stingrayb, huppenkothen_2019_stingraya}. The selected energy range is 1--10\,keV for LE, 10--30\,keV for ME, and 30--150\,keV for HE. In order to study the energy-dependent properties, the energy range of LE has been divided into 4 bands: 1--3\,keV, 3--5\,keV, 5--7\,keV, and 7--10\,keV; ME is divided into 2 bands: 10--20\,keV and 20--30\,keV; HE is divided into 3 bands: 30--50\,keV, 50--90\,keV, and 90--150\,keV. In the following, whenever hardnesses are discussed, they are calculated using raw photon counts, not fluxes. The time resolution of those light curves is set to be $2^{-9}$\,s $\approx 2$\,ms. As we are using the Fast Fourier Transform for our calculations, we use segments with a length of $2^{13} \times \Delta t = 16$\,s to calculate timing properties. The dead time of the LE telescope is negligible, and set to zero~\citep{chen_2020_low}; the dead time of ME and HE are set to 200\,$\mu$s and 2\,$\mu$s, respectively~\citep{cao_2020_medium, liu_2020_high}. The power spectral densities (PSDs) are Poisson-noise and dead-time corrected according to \citet{zhang_1995}. 

Thus, the Nyquist frequency for the light curves is $f_{\text{max}} = 1/(2\Delta t) = 256$\,Hz, and the lowest frequency we can access is $f_{\text{min}} = 1/(n_{\text{bins}} \Delta t) = 0.0625$\,Hz. 

\subsection{Long-term source behavior}\label{subsect:obs-longterm}

Different from the majority of the transient BHBs that trace the canonical \texttt{q}-track during the outburst, \cyg is one of the only few known persistent BHBs, fed by the wind from its companion. It has been bright since its discovery, and often moves between the hard state and the soft state, without tracing any \texttt{q}-track~\citep[e.g.,][]{wilms_2006, fender_2006, grinberg_2013}. Previous studies have shown that \cyg spends most of the time in the hard state and the soft state, which indicates that these two states are more stable than the intermediate state~\citep[see e.g.,][]{grinberg_2013, meyer-hofmeister_2020}. Prior to $\sim$ 2011, the source has spent most of the time in the hard state, with only occasional excursions towards the soft~\citep{grinberg_2013, grinberg_2014}. Since then and within the life time of MAXI, it has mainly been in the soft state~\citep{Cangemi_2021a}; the reason for this is not yet known. 

\begin{figure}
    \centering
    \includegraphics[width = 0.49\textwidth]{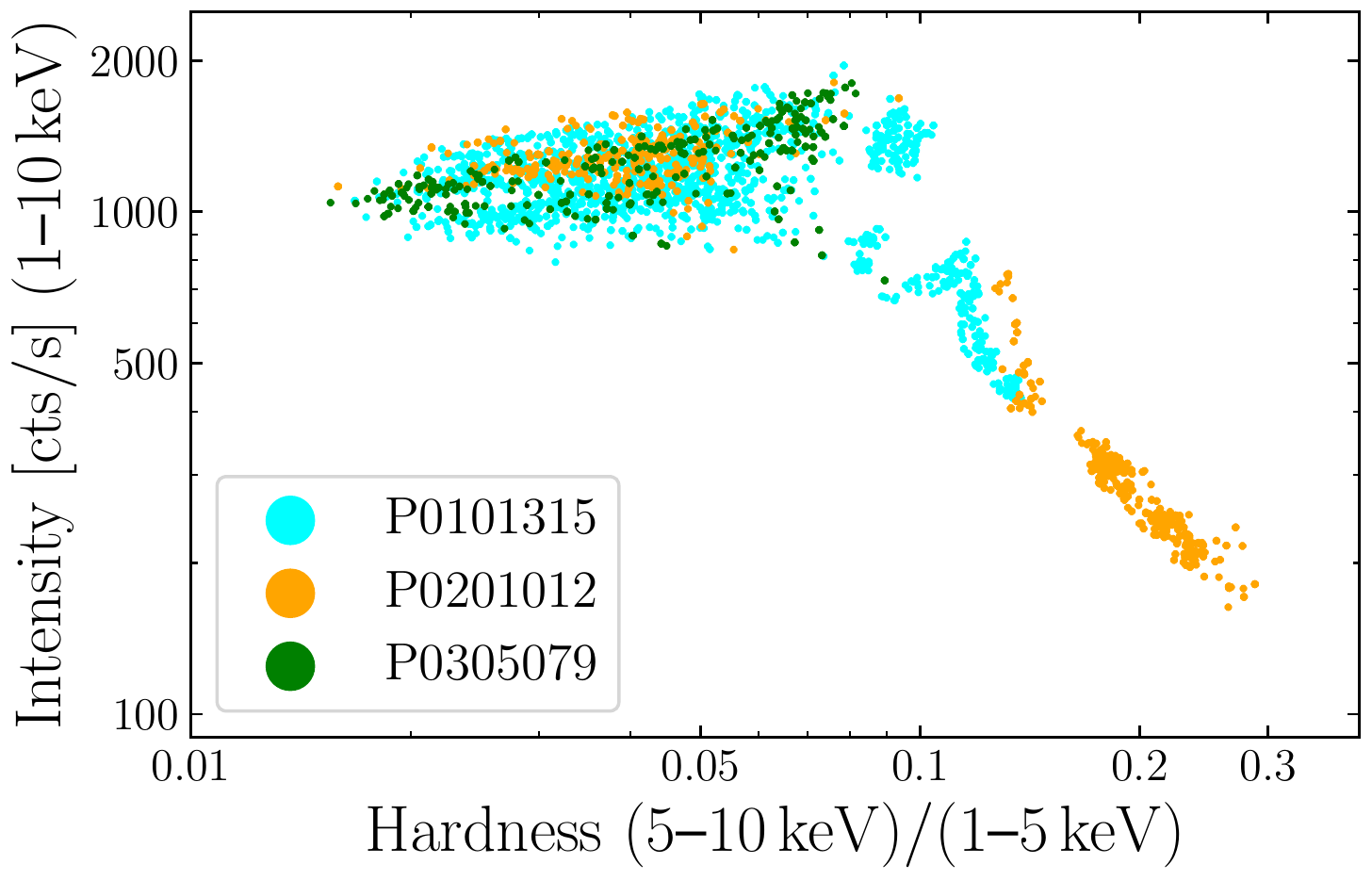}
    \caption{HID of \cyg by the data of three proposals: P0101315, P0201012, and P0305079, that are denoted by cyan, orange, and green, respectively. The intensity is defined by the photon count rate of LE ranging from 1 to 10\,keV. We define the hardness as the ratio of the raw photon count rate between 5--10\,keV band and 1--5\,keV band. The time resolution is set to be 60 seconds. }
    \label{f-HID}
\end{figure}

\citet{pottschmidt_2003} report on ``failed transitions'' frequently observed in \cyg: the source softens, but never reaches a proper soft state, instead returning to the hard state eventually. Similar failed transitions have also been observed in transient BHBs (e.g., GX~339$-$4, see \citealt{fuerst_2015}, \citealt{garcia_2019}; and Swift~J1753.5$-$0127, see \citealt{bu_2019}). Although enormous data have been collected over decades, the true mechanism of state transitions and the reason why BHBs maintain their stability particularly in the hard/soft state are still waiting to be solved. 

Here we use the LE data of the three proposals in Table~\ref{tab:proposal-id} to produce the HID for \cyg, as shown in Fig.~\ref{f-HID}. The data from proposal P0101315 and P0305079 cover the soft state and the intermediate state. However, the proposal P0201012 contains observations in the hard state, enabling us to compare source properties across all states. 

\section{Spectral analysis}\label{sect:spec}

\subsection{Spectral modeling}

\begin{figure}
    \centering
    \includegraphics[width = 0.49\textwidth]{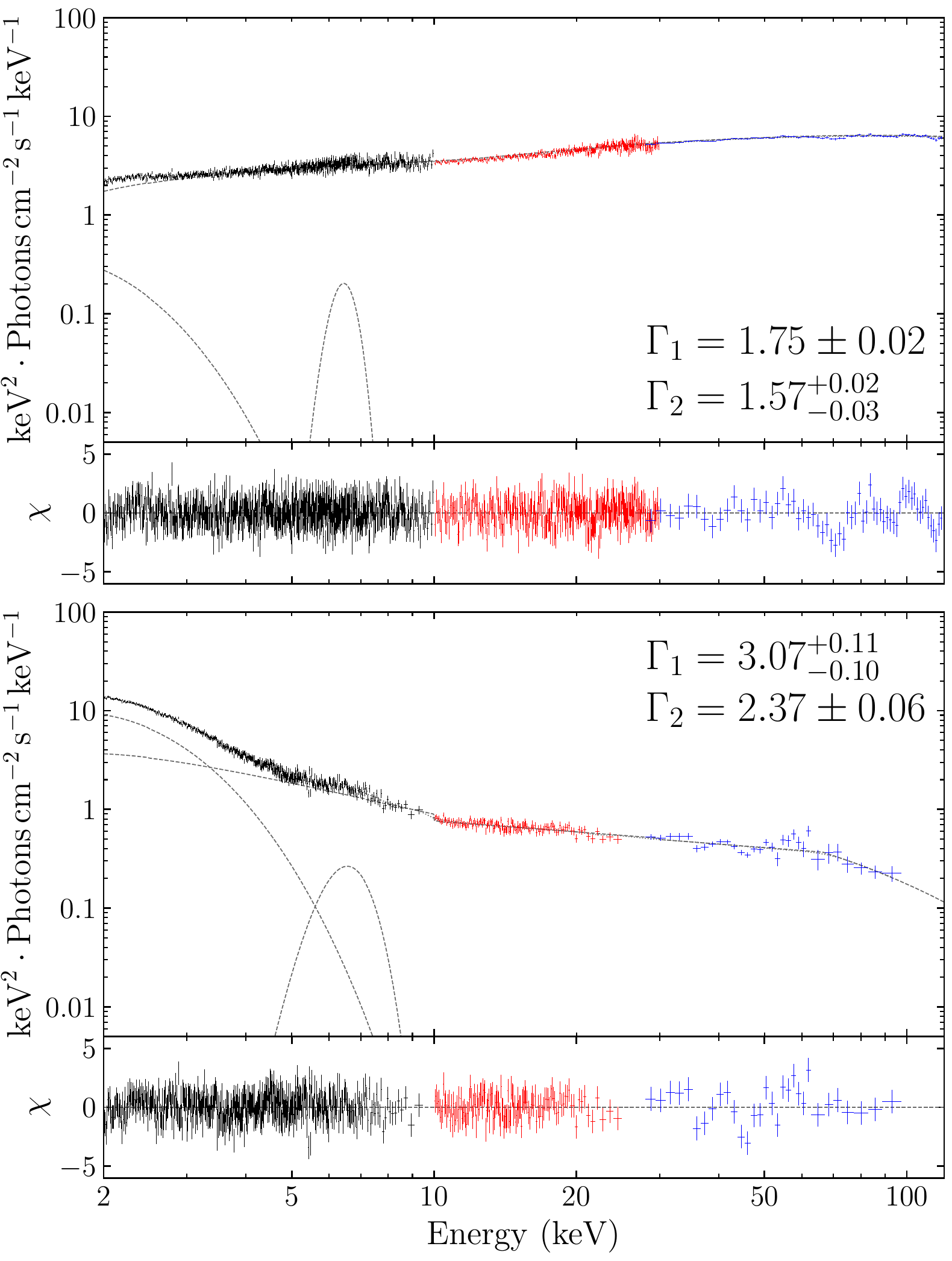}
    \caption{Typical spectra of \cyg with HXMT observations. The \textit{upper panel} shows a spectrum in the hard state (obs ID: P0201012171). The spectrum in the \textit{lower panel} is in the soft state (obs ID: P0101315057). Data from LE/ME/HE telescope are denoted with black/red/blue crosses. The powerlaw continuum, the thermal component, and the iron line emission are indicated with dashed lines. }
    \label{f-typicalspectra}
\end{figure}

For a spectral-timing study of numerous observation samples presented here, a reliable but simple characterization of the spectral shape is necessary. Former long-term spectral studies of \cyg show the advantages to use a simple phenomenological model consisting of a broken powerlaw modified by a high-energy cut-off and a disk component to fit the energy spectra~\citep{wilms_2006, boeck_2011, grinberg_2014}. We can write the model as: 
\begin{equation}
    \texttt{const} \times \texttt{tbabs} \times (\texttt{bknpower} \times \texttt{highecut} + \texttt{gaussian} + \texttt{diskbb}) \, , \notag
\end{equation}
where $\texttt{tbabs}$ calculates the photons absorbed by the interstellar medium, whose cross-sections are provided by \citet{verner_1996} and the element abundances are given by \citet{wilms_2000}. 

$\texttt{Bknpower} \times \texttt{highecut}$ empirically describes a continuum which dominates the spectrum in the hard X-ray band. $\texttt{Bknpower}$ has three parameters except for its normalization: $\Gamma_1$, the photon index of the soft band; $\Gamma_2$, the index of the hard band, and the breaking energy $E_{\text{break}}$, in our case around 10\,keV. We also included in the fit a Gaussian line to characterize the iron K$\alpha$ line emission located at around 6.4\,keV. Finally, we used $\texttt{diskbb}$, a multi-temperature disk model, to fit the thermal radiation from the accretion disk~\citep{mitsuda_1984, makishima_1986}. Typical spectra for \cyg are shown in Fig.~\ref{f-typicalspectra}. In the hard state, the disk component never dominates the spectrum. However, the thermal component plays a significant role at soft X-rays when the source is in the soft state. 

For our study, this simple phenomenological modelization is sufficient to identify the different spectral component, but we note here that it would also be worthwhile to use a more sophisticated physical model to fit the spectra, involving more parameters and more complex assumptions~\citep[for a recent study using a physical model on \cyg with the HXMT data, see e.g.,][]{feng_2022}. 

The data coming from all instruments are rebinned to a minimum signal-to-noise ratio of 10. 
Considering the accuracy of the current calibration and that \cyg is a bright source, we adopt a systematic error of 0.5\%, 0.5\%, and 1.0\% for LE, ME, and HE, respectively~\citep{li_2020_flight, kong_2021}. The peak energy of $\texttt{gaussian}$ is fixed at 6.4\,keV, and the column density $N_\mathrm{H}$ is fixed at the value of $7.1 \times {10}^{21} \text{cm}^{-2}$, as given by the measurements of the HI4PI survey based on EBHIS and GASS~\citep{bekhti_2016}. All the other parameters remain free. 

\begin{figure}
    \centering
    \includegraphics[width = 0.49\textwidth]{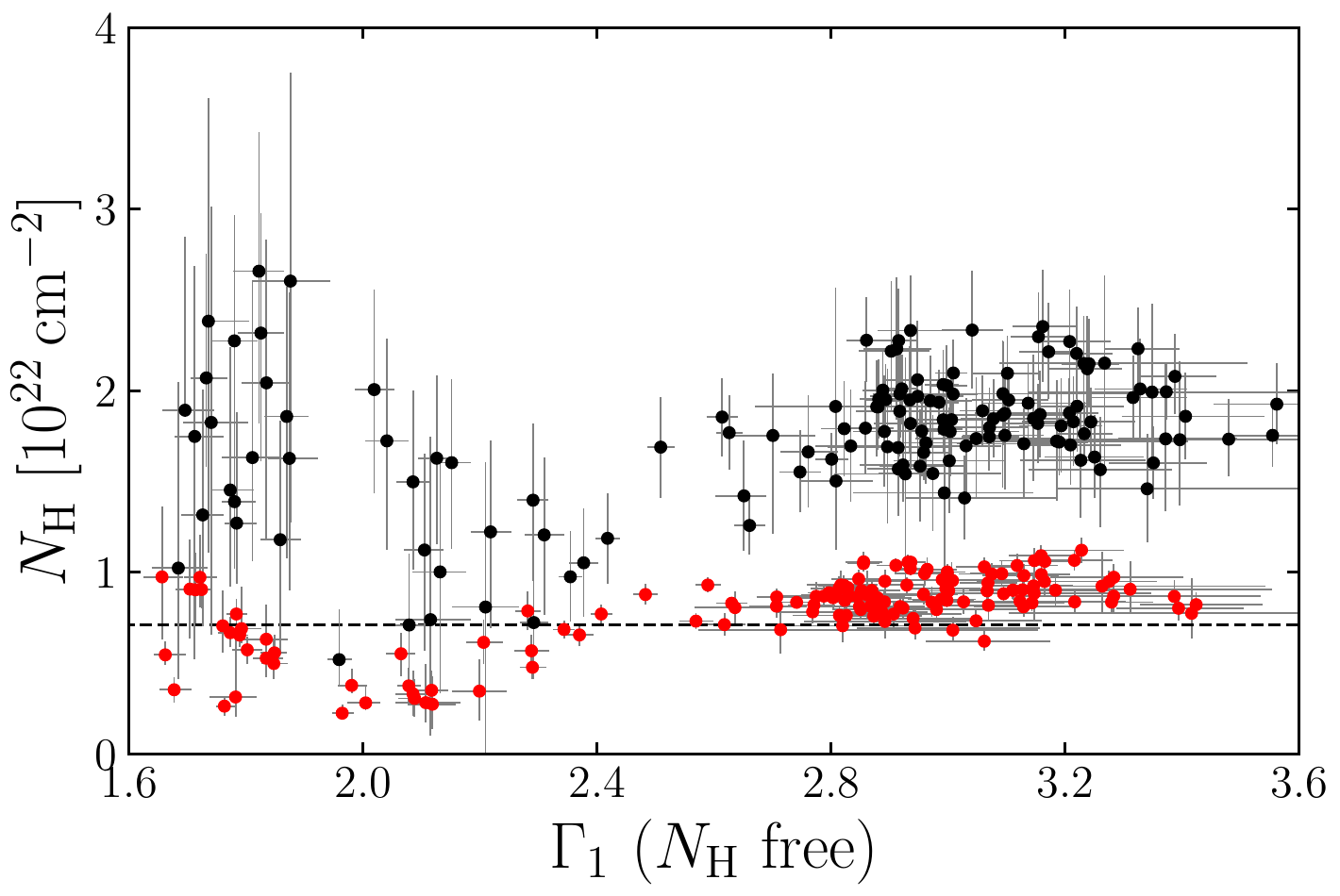}
    \caption{The soft photon index $\Gamma_{1}$ of \texttt{bknpower} obtained by free column densities versus the obtained column densities $N_\mathrm{H}$ during the absorption. The black dots indicate the $N_\mathrm{H}$ obtained by the spectral fitting down to 2\,keV; while the red dots indicate the $N_\mathrm{H}$ obtained down to 1.5\,keV. The dashed line represents the value $7.1 \times {10}^{21} \text{cm}^{-2}$, which is adopted during our spectral fitting. }
    \label{f-index1vsnH}
\end{figure}

\begin{figure}
    \centering
    \includegraphics[width = 0.49\textwidth]{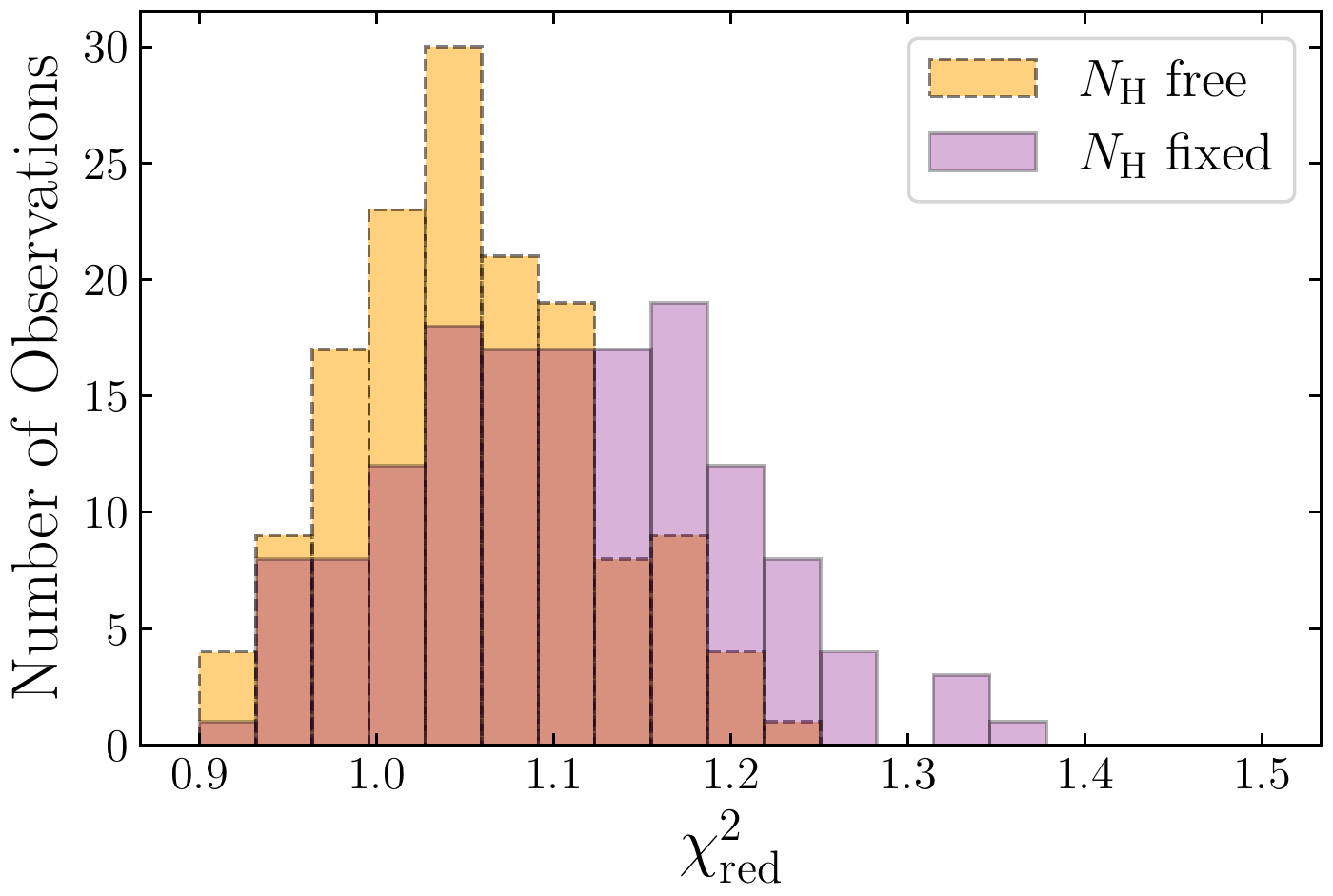}
    \caption{The histogram of the reduced $\chi^2$ over the best fits of the 145 spectra. Most of the $\chi^2_{\text{red}}$ with fixed column densities at $7.1 \times {10}^{21}\,\text{cm}^{-2}$ (in translucent purple) drop in the range 0.9 to 1.2, which suggests that the best fit is relatively satisfying. The bars in orange are the distribution of $\chi^2_{\text{red}}$ with free column densities. }
    \label{f-hist-rchi2}
\end{figure}

While we in general expect $N_\mathrm{H}$ to be variable in \cyg during the hard state especially \citep[e.g.,][]{grinberg_2015,Hirsch_2019a,Lai_2022a}, only a handful of our observations are in the hard state at the orbital phases where they could be affected. In the soft state, i.e., the state of most of our observations, the wind is expected to be highly ionized and thus would contribute little to the observed absorption. The $N_\mathrm{H}$ measurements obtained from our HXMT data if the absorption column density is left free to vary, show trends that we interpret as systematic effects. In particular, in Fig.~\ref{f-index1vsnH} (in black), we can see a clear increase in $N_\mathrm{H}$ for higher $\Gamma_1$, which we interpret as due to the strong contribution of the steep power law at low energy and the degeneracies between the power law, disk, and absorption contributions. This is particularly the case when the power law is steep and the contribution of the disk is strong, because the simple power law model, which does not have a low energy cut-off, extends below the energies of the putative seed photon contribution (i.e., the \texttt{diskbb} component) for the Comptonization component it is describing. Similar overestimation of $N_\mathrm{H}$ can also be seen in, e.g., \citet{boeck_2011} and \citet{grinberg_2015}, where the high $N_\mathrm{H}$  are associated with high uncertainties.
We conduct an additional test by extending the lower limit of our spectra to  1.5\,keV, taking advantage of the LE instrument response. We obtain better constraints on $N_\mathrm{H}$ that are very close to the value suggested by the HI4PI survey (Fig.~\ref{f-index1vsnH}, in red). Since this work does not focus on the contribution of the stellar wind but aims to obtain a satisfactory continuum description, we interpret this as confirming our choice to fix the $N_\mathrm{H}$ value, but caution that detailed disk and wind studies would require a careful consideration of possible calibration effect and model degeneracies. 

Overall, we are able to obtain good fits for all our spectra as shown in Fig.~\ref{f-hist-rchi2}, where we also show the quality of fits obtained when the column density is left free.

\subsection{Discussion of spectral modeling}\label{subsect:spec-discussion}

\begin{figure}
    \centering
    \includegraphics[width = 0.49\textwidth]{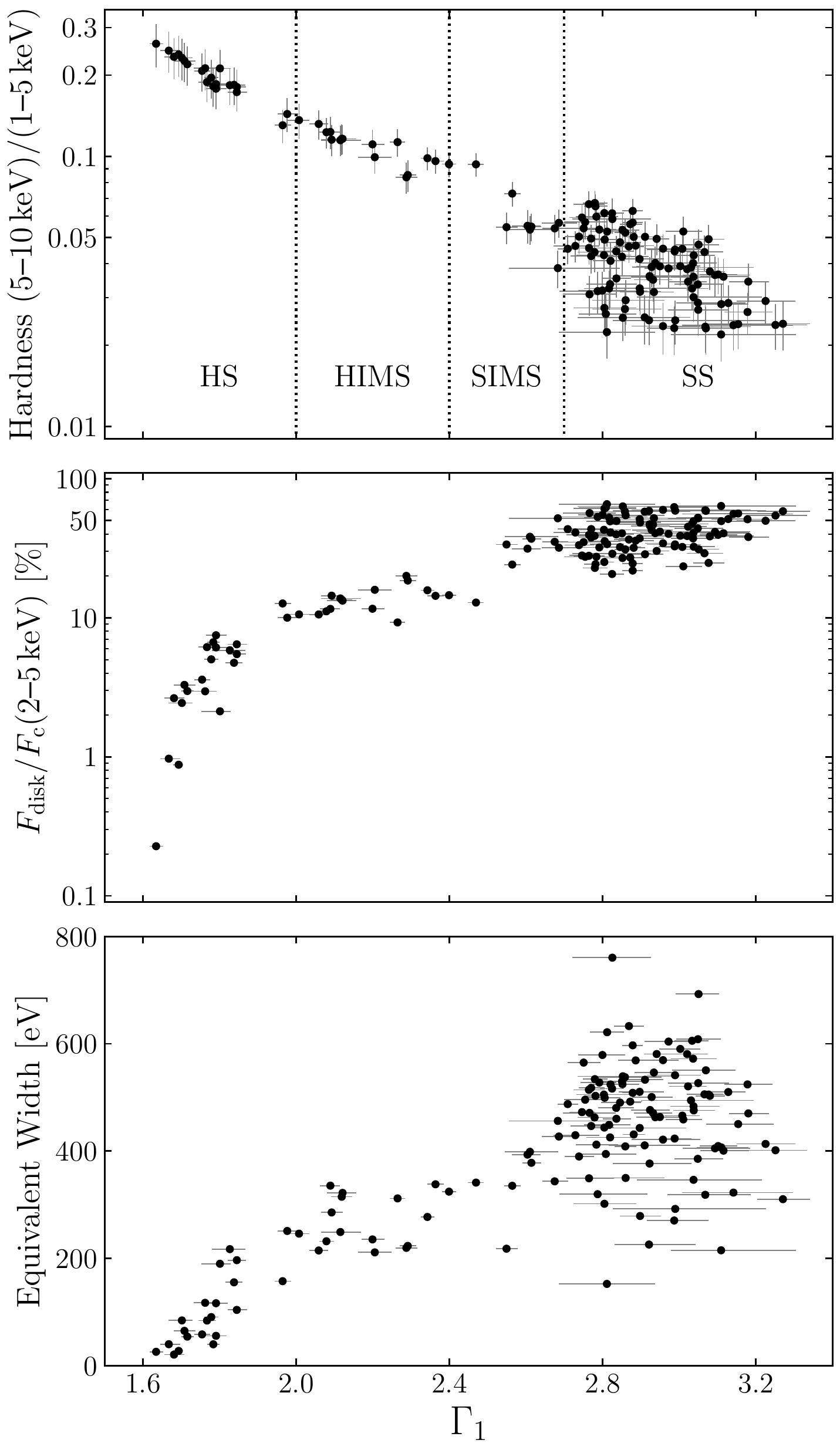}
    \caption{The \textit{top panel} shows the correlations of the 
    5--10\,keV / 1--5\,keV hardness versus the soft photon index $\Gamma_1$. We thus define the hard state (HS), the hard intermediate state (HIMS), the soft intermediate state (SIMS), and the soft state (SS) with $\Gamma_1$ values. The \textit{middle panel} shows the ratio of unabsorbed energy flux of the disk components over the total energy flux versus $\Gamma_1$. The \textit{bottom panel} shows the intensity of reflection that is denoted by the equivalent widths of the iron K$\alpha$ lines, is stronger in the soft state than in the harder state. }
    \label{f-hdvsmodelflux}
\end{figure}

We first verify the validity of our spectral modeling by checking the correlations between the spectral hardness and soft photon index $\Gamma_1$, and between the ratio of unabsorbed energy flux directly from the thermal radiation (\texttt{diskbb}-model) over the total unabsorbed energy flux in the 2--5\,keV range and $\Gamma_1$, as shown in the first two panels in Fig.~\ref{f-hdvsmodelflux}. Both show the correlations expected for hard to soft transition. 

We now can define the state of the source by the corresponding value of $\Gamma_1$: the hard state for $\Gamma_1 \leq 2.0$, the hard intermediate state for $2.0 < \Gamma_1 \leq 2.4$, the soft intermediate state for $2.4 < \Gamma_1 \leq 2.7$, and the soft state if $\Gamma_1 > 2.7$. Our choice is supported by the sudden change of the PSD shape in the 3--5\,keV band and the intrinsic coherence in the 1--3 vs. 3--5\,keV band, see Sect.~\ref{subsect:discussion_1-PSDevo} and~\ref{subsect:discussion_1-coh-phlag} for more details as well as \citet{grinberg_2014}. 

The positive correlation between the intensity of the reflection, shown by the equivalent width of the iron K$\alpha$ line, and the spectral parameter $\Gamma_1$, is presented in the bottom panel of Fig.~\ref{f-hdvsmodelflux}. This result is consistent with e.g., \citet{gilfanov_1999}, \citet{ibragimov_2005}, \citet{shaposhnikov_2006}, and \citet{steiner_2016}, who all used \textsl{RXTE} data for their analyses. 

By studying \cyg, \citet{shaposhnikov_2006} propose that the corona may have a compact geometry in the intermediate and soft state. Based on observations of 29 stellar-mass BH candidates, \citet{steiner_2016} offer two interpretations. One agrees with \citet{shaposhnikov_2006} that the size of the corona shrinks when the state softens, and the other agrees with \citet{petrucci_2001}, who discuss the Comptonization effect on the reflection component, stating that in the hard state, the iron line amplitude will be diluted by Compton scattering due to the higher optical depth of the corona. 

\begin{figure}
    \centering
    \includegraphics[width = 0.49\textwidth]{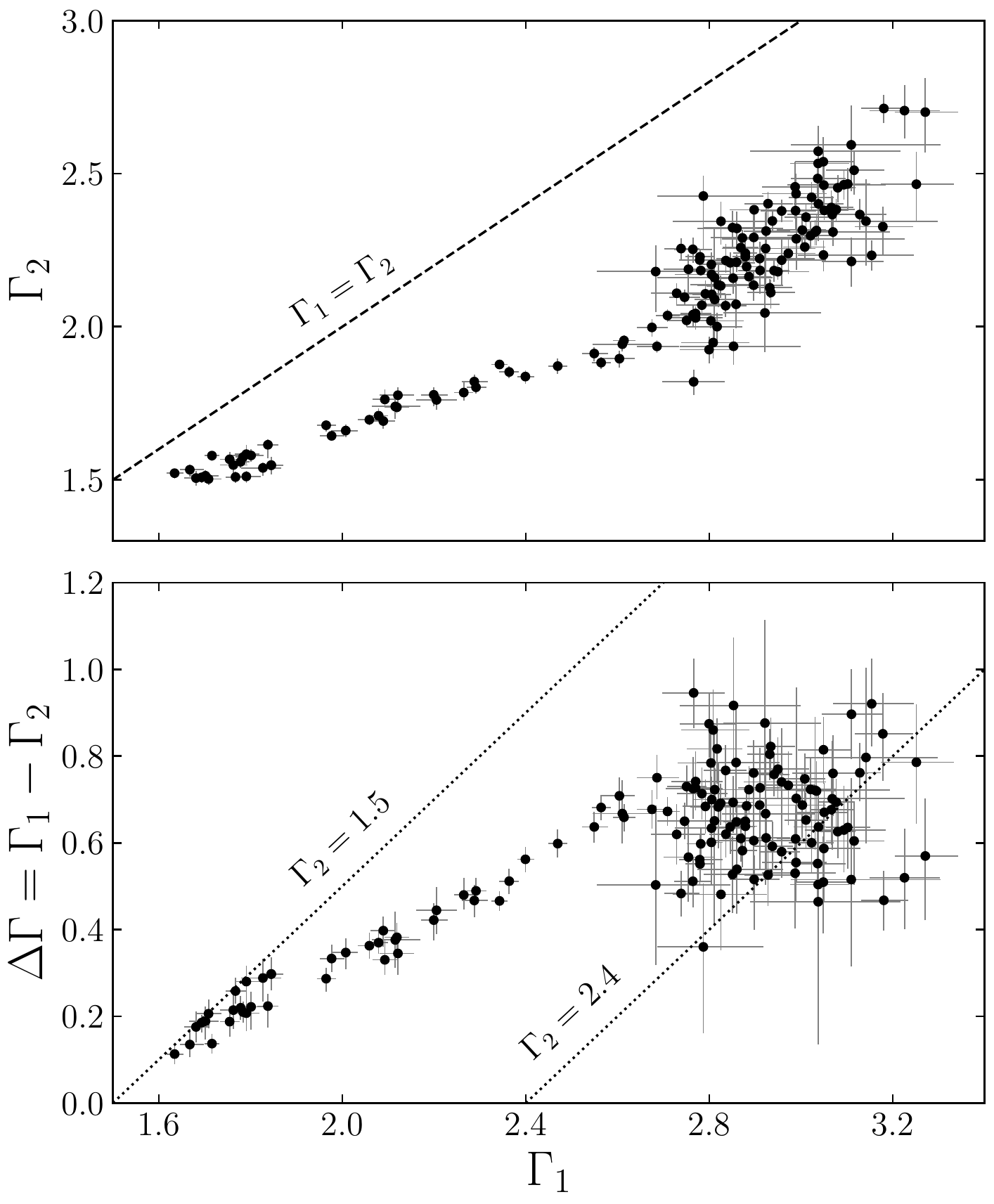}
    \caption{The \textit{upper panel} shows the soft photon index $\Gamma_{1}$ versus the hard photon index $\Gamma_{2}$ in model \texttt{bknpower}. The dashed line indicates that $\Gamma_{1} = \Gamma_{2}$. In the \textit{lower panel}, the correlations between $\Gamma_{1}$ and the difference of two photon indices, defined by $\Delta\Gamma = \Gamma_{1} - \Gamma_{2}$, are shown. The lines that are parallel to the dotted lines indicate a constant $\Gamma_{2}$. }
    \label{f-index1vs2}
\end{figure}

\begin{figure}
    \centering
    \includegraphics[width = 0.49\textwidth]{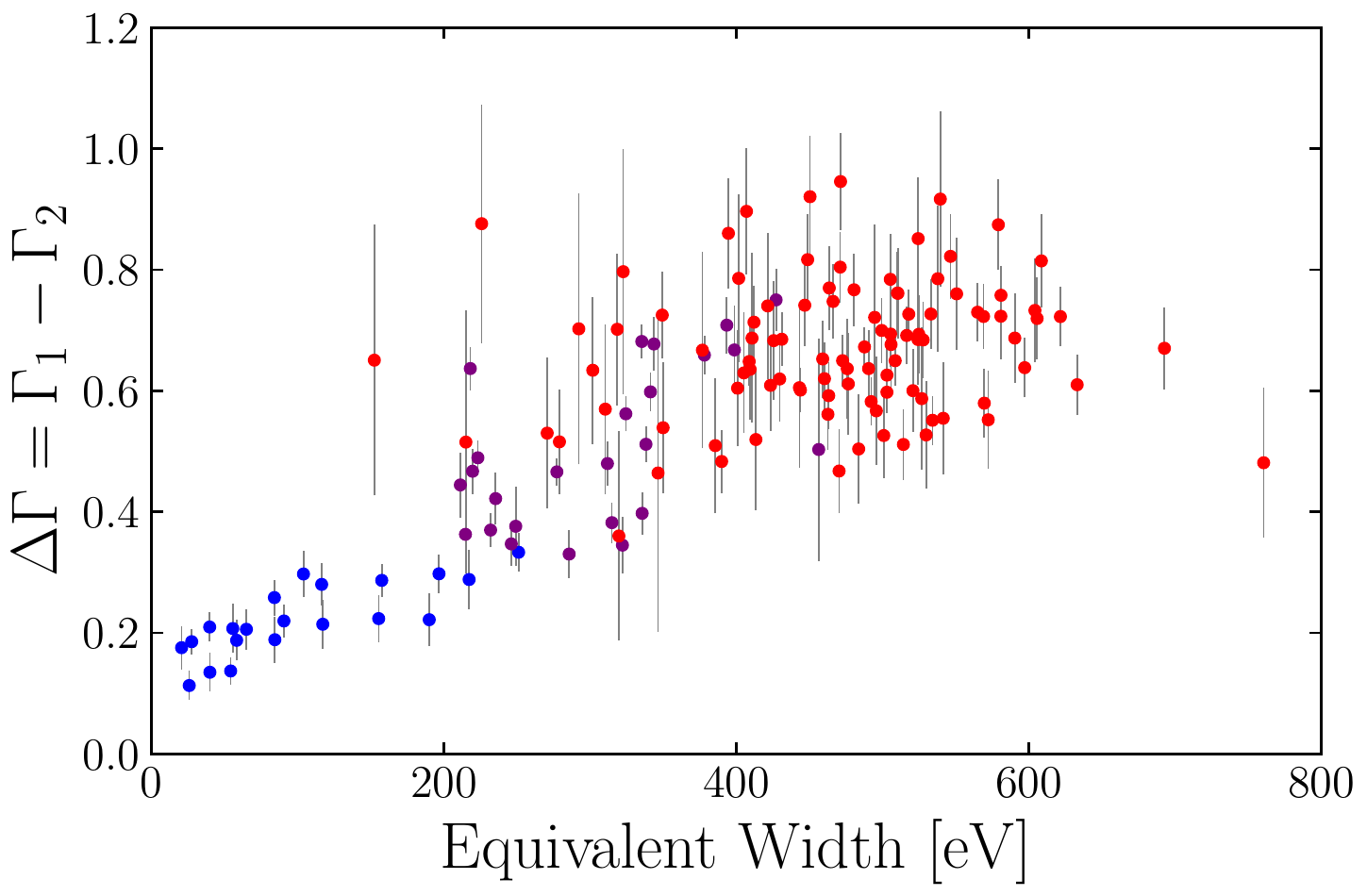}
    \caption{The correlation of $\Delta\Gamma$ versus the equivalent widths of the iron K$\alpha$ lines. The blue/purple/red points denote that the source is in the hard/intermediate/soft state, respectively. }
    \label{f-eqvsdelgamma}
\end{figure}

The intensity of the reflection can be roughly quantified by the difference of the two photon indices as well, as the model combination $\texttt{bknpower} \times \texttt{highecut}$ with the breaking energy around 10\,keV is capable to characterize the Compton hump around 20--30\,keV~\citep{wilms_2006}. We present the correlations of two photon indices and their differences $\Delta \Gamma$ in Fig.~\ref{f-index1vs2}, and $\Delta \Gamma$ versus the equivalent widths of the iron lines in Fig.~\ref{f-eqvsdelgamma}. Consistent with the tendency between the equivalent width and $\Gamma_1$, we see a clear positive trend between $\Delta \Gamma$ and $\Gamma_1$. In addition, Fig.~\ref{f-eqvsdelgamma} shows that the two indicators of the reflectional strength are indeed positively correlated even in the soft state. We thus conclude that in the soft state, the reflection is indeed stronger than that in the hard state. 

\begin{figure}
    \centering
    \includegraphics[width = 0.49\textwidth]{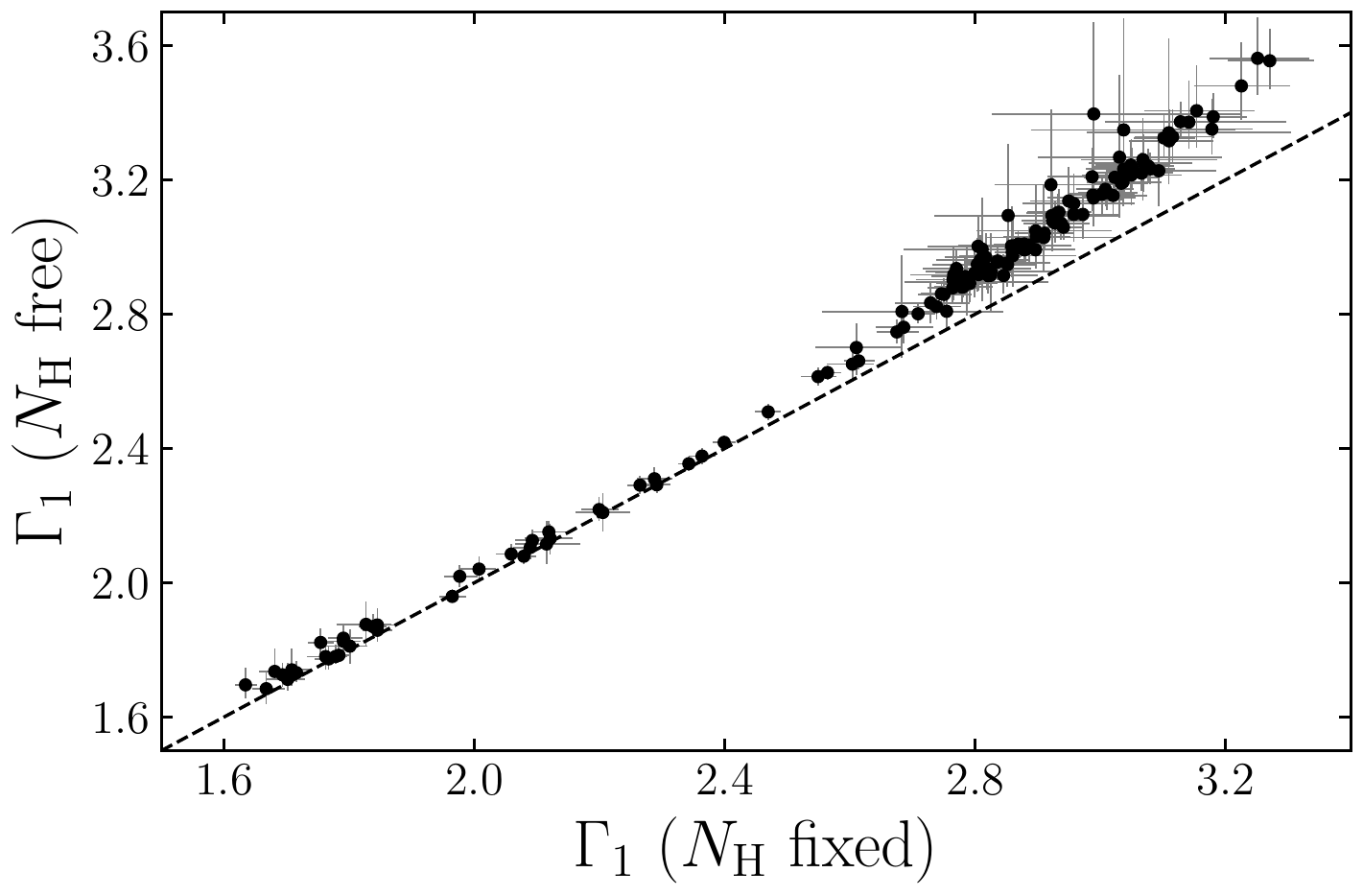}
    \caption{The soft photon index $\Gamma_{1}$ of \texttt{bknpower} obtained by fixed $N_\mathrm{H} = 7.1 \times {10}^{21} \text{cm}^{-2}$ and free $N_\mathrm{H}$. The dashed line indicates that $\Gamma_{1} \  (N_\mathrm{H}\ \text{free}) = \Gamma_{1} \  (N_\mathrm{H}\ \text{fixed})$. }
    \label{f-index1vs1}
\end{figure}

However, the correlation we observe between $\Gamma_{1}$ and $\Gamma_{2}$ shows a ``bend'' at $\Gamma_1 \approx 2.7$ as shown in the upper panel of Fig.~\ref{f-index1vs2}. This is coincident with the onset of the soft state and fits where the uncertainties on the current steep power law index are larger. Similar trends were not observed in \citet{wilms_2006}, who did not include a disk for their fits in the soft state, resulting in a bad description of the spectra below 6\,keV. We check whether our choice to freeze $N_\mathrm{H}$ may lead to the observed bend in the correlation:
We can find that for $\Gamma_1 \ (N_\mathrm{H}\ \text{fixed}) \gtrsim 2.7$, an evident deviation between the $\Gamma_1$s with fixed and free $N_\mathrm{H}$ appear (see Fig.~\ref{f-index1vs1}). Still, even for free $N_\mathrm{H}$ the bend in the correlation appears, i.e., it is not purely an artifact of our choice of parameter space of our model. 

\section{Energy-resolved variability properties}\label{sect:timing}
\subsection{Power spectra and the fractional rms}\label{subsect:timing-rms}

Power spectra (also power spectral densities, or PSDs), are the first order products of the Discrete Fourier Transform performed on a time series, and their calculations are discussed in detail by e.g., Sect.~2 of \citet{nowak_1999}. The power spectra are calculated in units of rms fluctuations scaled to the mean count rate~\citep{belloni_1990_variability, miyamoto_1991}. This normalization is often employed, as it is convenient to compare the power spectra from different states or even different sources, because the variance that each frequency contributes is independent of the averaged photon count rate. The total fractional rms is the integral of the PSDs over all available frequencies, which in our case is $0.0625\text{--}256\,$Hz. 

In some of the energy bands considered for our analysis, the background photons contribute significantly. Especially in the soft state, they dominate in the hard X-ray bands and need to be taken into account. We thus adopt a background correction for the fractional rms \citep{belloni_1990_atlas, bu_2015}:
\begin{equation}\label{Bcorrection}
    \text{rms} =  \sqrt{\sum P \cdot \Delta f} \  \left( \frac{S + B}{S} \right) \, ,
\end{equation}
where $P$ is the noise-subtracted power normalized according to \citet{miyamoto_1991}, and $S$ and $B$ are the averaged photon count rates from the source and the background. This background correction adds a factor $(S + B)/S$ which is always greater than 1 to the raw rms values. It assumes that the background is intrinsically invariant and therefore does not contribute to the total variability. As expected, this assumption is not true when background noise dominates the data, and it will cause the overestimation of the fractional rms, since the background noise has its own intrinsic variability as well.

\begin{figure*}
    \centering
    \includegraphics[width = 0.99\textwidth]{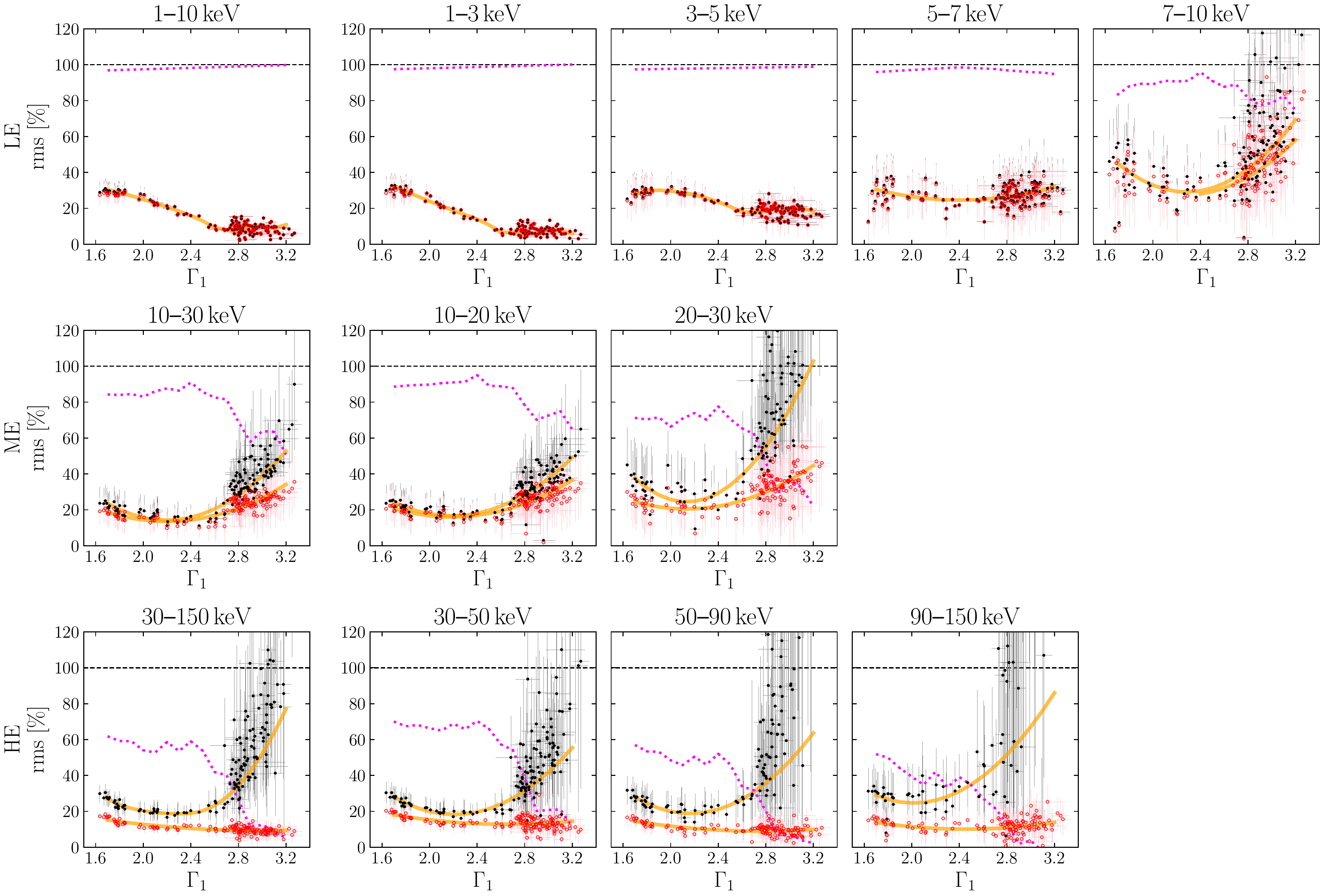}
    \caption{The correlations between the soft photon index $\Gamma_1$ and the fractional rms from all the instruments on HXMT. The red circles are the raw rms without the correction of Eq.~\ref{Bcorrection}, and the black dots are the corrected total rms according to Eq.~\ref{Bcorrection}. The magenta dotted lines denote the signal-background ratio $S/(S + B)$. The orange solid lines show the trends of the corrected/uncorrected total rms when varying $\Gamma_1$. } 
    \label{f-index1vsrms}
\end{figure*}

In Fig.~\ref{f-index1vsrms} we show the correlations of different estimates of total rms with spectral shape as represented by $\Gamma_1$ for all instruments and energy ranges under consideration. To present the effect of the employed background correction, we additionally show the inverse of the correction factor, $[(S + B)/S]^{-1} = S/(S + B)$. For $S/(S + B)$ close to 1, most of the photons come from the source, and therefore the corrected rms is almost equal to the raw rms. The corrected rms is higher than the raw rms when $S/(S + B) < 1$. Since some of the data with a high $\Gamma_1$ are very noisy, we use smooth curves calculated through simple interpolation between data points to present the macroscopic trends between the total rms and $\Gamma_1$. 

\subsection{PSD evolution with spectral shape}\label{subsect:timing-PSDevo}

\begin{figure*}
    \centering
    \includegraphics[width = 0.99\textwidth]{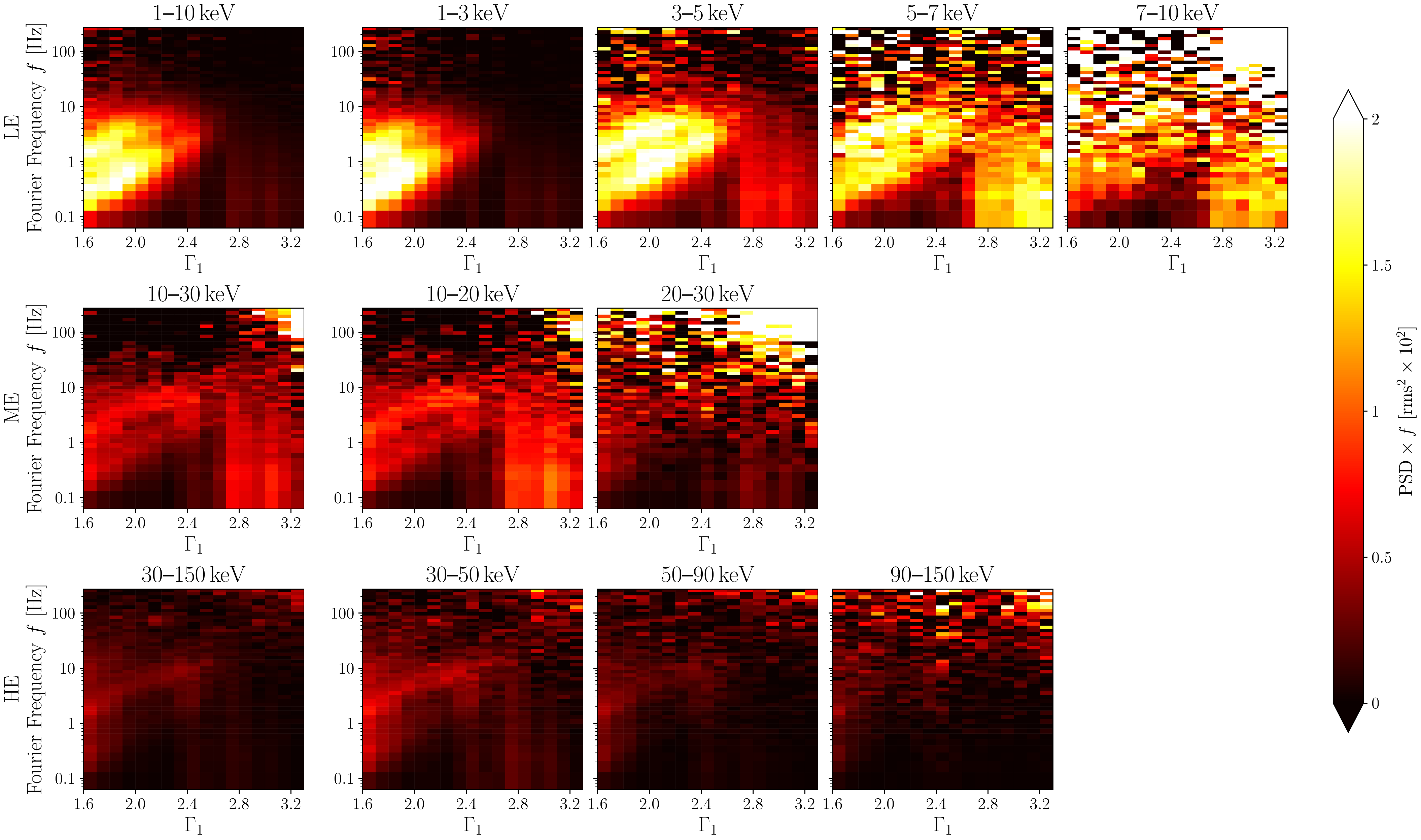}
    \caption{The evolution of PSDs versus the soft photon index $\Gamma_1$ from different energy bands provided by all instruments on HXMT. The color scale represents the averaged value of $\text{PSD}(f_i) \times f_i$ at individual frequencies $f_i$. Please note that the PSD values in these plots have not been corrected by Eq.~\ref{Bcorrection} in order to avoid the covering effect due to high signal-background ratio. } 
    \label{f-PSD-evo}
\end{figure*}

We follow the approach of \citet{boeck_2011} and \citet{grinberg_2014} to visualize the whole power spectra by color maps in $\Gamma_1$-$f_i$-space. The values of $\Gamma_1$ are obtained from the best fits from Sect.~\ref{sect:spec}, where $\Gamma_{1,\, \mathrm{min}} = 1.63$ and $\Gamma_{1,\, \mathrm{max}} = 3.27$. The grid of $\Gamma_1$ is defined linearly by constant steps of $\Delta\Gamma_1 = 0.1$ between the two limits. The power spectra from different observations will be averaged if their $\Gamma_1$ belongs to the same bin of the grid. The frequencies $f$ are rebinned logarithmically between 0.0625\,Hz and 256\,Hz, with an equally spaced grid in logarithmic scale $\mathrm{d}f/f = 0.15$. The evolution of the PSD shape versus $\Gamma_1$ is shown in Fig.~\ref{f-PSD-evo}. 

\begin{figure*}
    \centering
    \includegraphics[width = 0.99\textwidth]{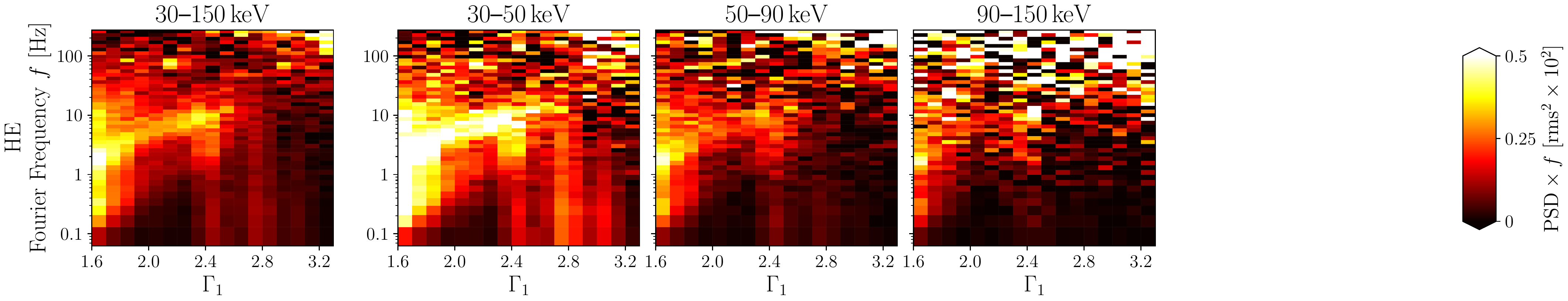}
    \caption{Same as the last row of Fig.~\ref{f-PSD-evo}, but with a different color scale, in order to illustrate the evolution of the variability component more clearly. } 
    \label{f-PSD-evo-he}
\end{figure*}

The PSDs shown in Fig.~\ref{f-PSD-evo} have not been corrected by Eq.~\ref{Bcorrection}, thus the integral over all frequencies corresponds to the uncorrected rms shown in Fig.~\ref{f-index1vsrms}. We can also see that the strength of the variability from HE bands is weaker than that from ME bands. The main reason is that the background photons that significantly contribute to the harder X-ray bands will elevate the detected count rate, hence the underestimation of the strength of the intrinsic variability of the source. To clearly show the evolution of the variability above 30\,keV, we re-scale the color map of HE bands, which shall be seen in Fig.~\ref{f-PSD-evo-he}. 

In Fig.~\ref{f-psd-le}, Fig.~\ref{f-psd-me} and Fig.~\ref{f-psd-he}, we show the averaged PSD shapes versus $\Gamma_1$ for all the three instruments. We note that ME and HE bands are both dominated by background photons when \cyg is in the extreme soft state. The flux in those bands are weak, and the uncertainties of the Fourier signals at high frequency bands are large. In this case, those PSDs should not be considered as physical. 

\subsection{Coherence function and phase-lags}\label{subsect:timing-coh-phlag}

The coherence, as well as the phase-lags, are the products of Fourier transform between two simultaneous light curve series. Usually we study these quantities between two different energy bands, to tell the lost phase-related information when we compute the moduli of the products of Fourier Transform. The intrinsic coherence function, $\gamma^2 (f)$, is defined by \citet{vaughan_1997}, where the auxiliary quantity $n^2$ defined in Eq.~4 by \citet{vaughan_1997} is the expectation of the noise contribution. 

The coherence function itself is a frequency-dependent measurement of the linear correlation between two simultaneous time series. In principle, the coherence function ranges from 0 to 1, where 0 means no coherence and 1 means perfect linear coherence. But when considering the noise correction, we possibly obtain a coherence greater than 1 or even negative. A negative coherence can merely emerge if the noise dominates one of the power spectra. In this case, the negative coherence there does not make sense. However, a coherence greater than 1 may be seen even when the signals at that frequency are well measured. This is due to the fact that the auxiliary quantity $n^2$ cannot be well-estimated by insufficient segments, or when the signals from two bands correlate extraordinarily strongly. Thus, the coherence higher than 1 still means good linear correlation between the two light curves and is not considered problematic in the following parts. 

\begin{figure*}
    \centering
    \includegraphics[width = 0.99\textwidth]{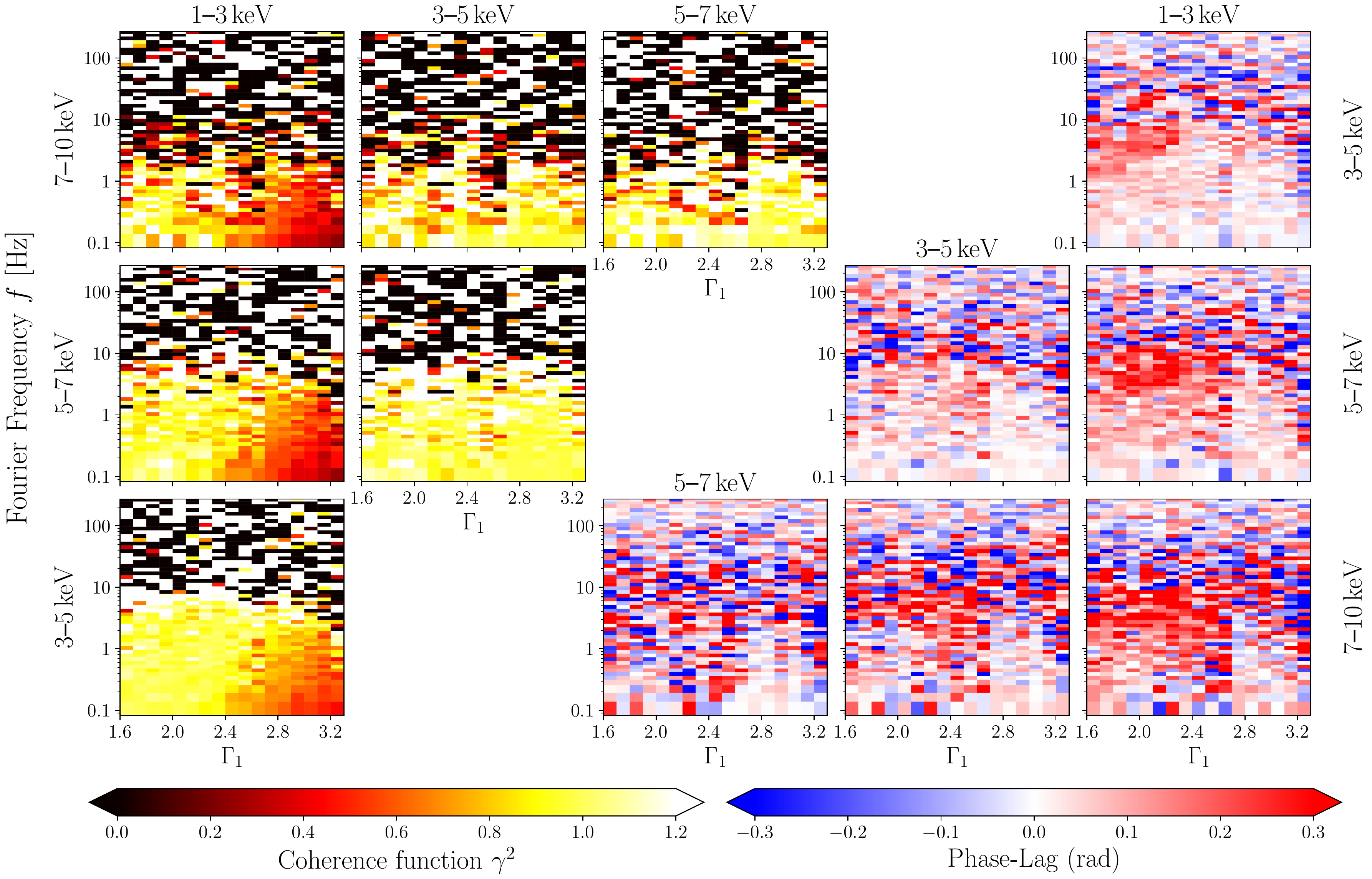}
    \caption{The evolution of the coherence function $\gamma^2$ and the phase-lags versus the soft photon index $\Gamma_1$ between different energy bands of the LE telescope. }
    \label{f-coh-lag-LE}
\end{figure*}

We follow the same recipe to present the coherence function in $\Gamma_1\text{-}f_i\text{-}\text{space}$, as shown on the left-hand side of Fig.~\ref{f-coh-lag-LE}. Similar as Fig.~\ref{f-PSD-evo} and Fig.~\ref{f-PSD-evo-he}, we use a thermal color map to visualize the coherence, where brighter colors represent better coherence, and dimmer colors represent worse coherence. We can see a clear envelope located at $\sim 10$\,Hz in each panel on the left-hand side of Fig.~\ref{f-coh-lag-LE}, and above $\sim 10$\,Hz, the noise-subtracted power densities are close to 0, and the coherence fluctuates violently then. 

The three panels in the first column of Fig.~\ref{f-coh-lag-LE} are the coherence functions of the softest band (1--3\,keV) versus the other harder bands (3--5\,keV, 5--7\,keV, and 7--10\,keV). In the hard state and the hard intermediate state, where $1.6 \leq \Gamma_1 \lesssim 2.4$, the coherence function in 1--3 vs. 3--5\,keV band is quite close to 1. But in the soft intermediate state and the soft state where $2.4 \lesssim \Gamma_1 \lesssim 3.3$, the loss of coherence emerges at low frequencies. Similar loss of coherence can also be found in 1--3 vs. 5--7\,keV band and 1--3 vs. 7--10\,keV band. The loss of coherence in the soft/intermediate state in these bands is due to the fact that the dominant spectral component in 1--3\,keV band has changed during the state transition. In the hard state where $\Gamma_1 \leq 2.0$, the 1--3\,keV band is dominated by the powerlaw emission. But in the soft state where $\Gamma_1 \gtrsim 2.7$, the thermal emission from the disk becomes dominant. As expected, and since the dominant emission comes from spatially distinct regions, we observe a loss of the coherence. 

When the two energy bands are both powerlaw dominated, the loss of coherence in the soft state disappears, which can be seen in the remaining three panels on the left-hand side of Fig.~\ref{f-coh-lag-LE}. We obtain best coherence for nearly all $\Gamma_1$ values in 3--5 vs. 5--7\,keV band. 

\begin{figure*}
    \sidecaption
    \includegraphics[width = 0.7\textwidth]{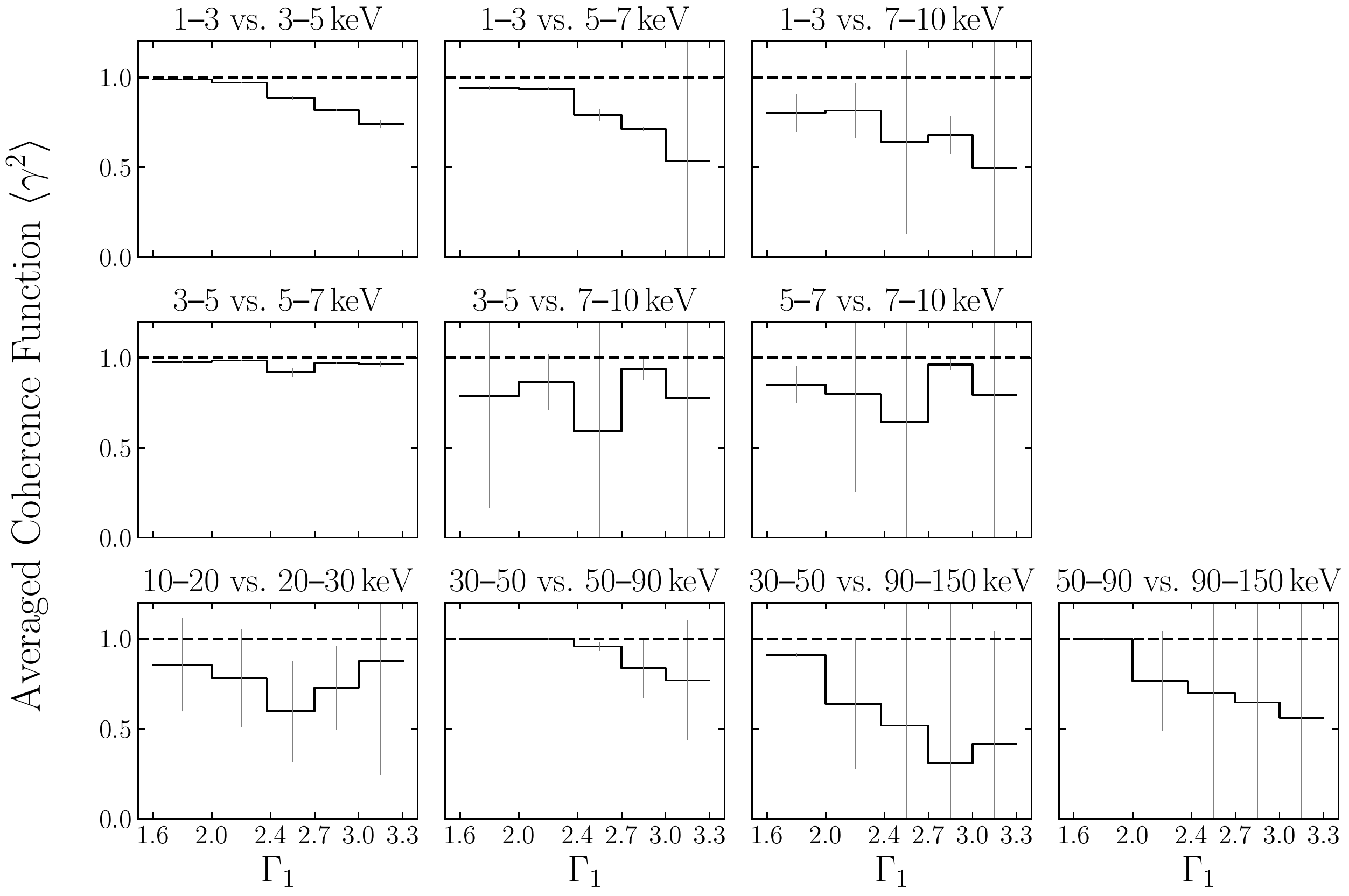}
    \caption{The evolution of the coherence $\langle \gamma^2 \rangle$ averaged in 0.0625--6\,Hz frequency range versus the states (classified by the soft photon index $\Gamma_1$) between different energy bands. }
    \label{f-coh}
\end{figure*}

\begin{figure*}
    \sidecaption
    \includegraphics[width = 0.7\textwidth]{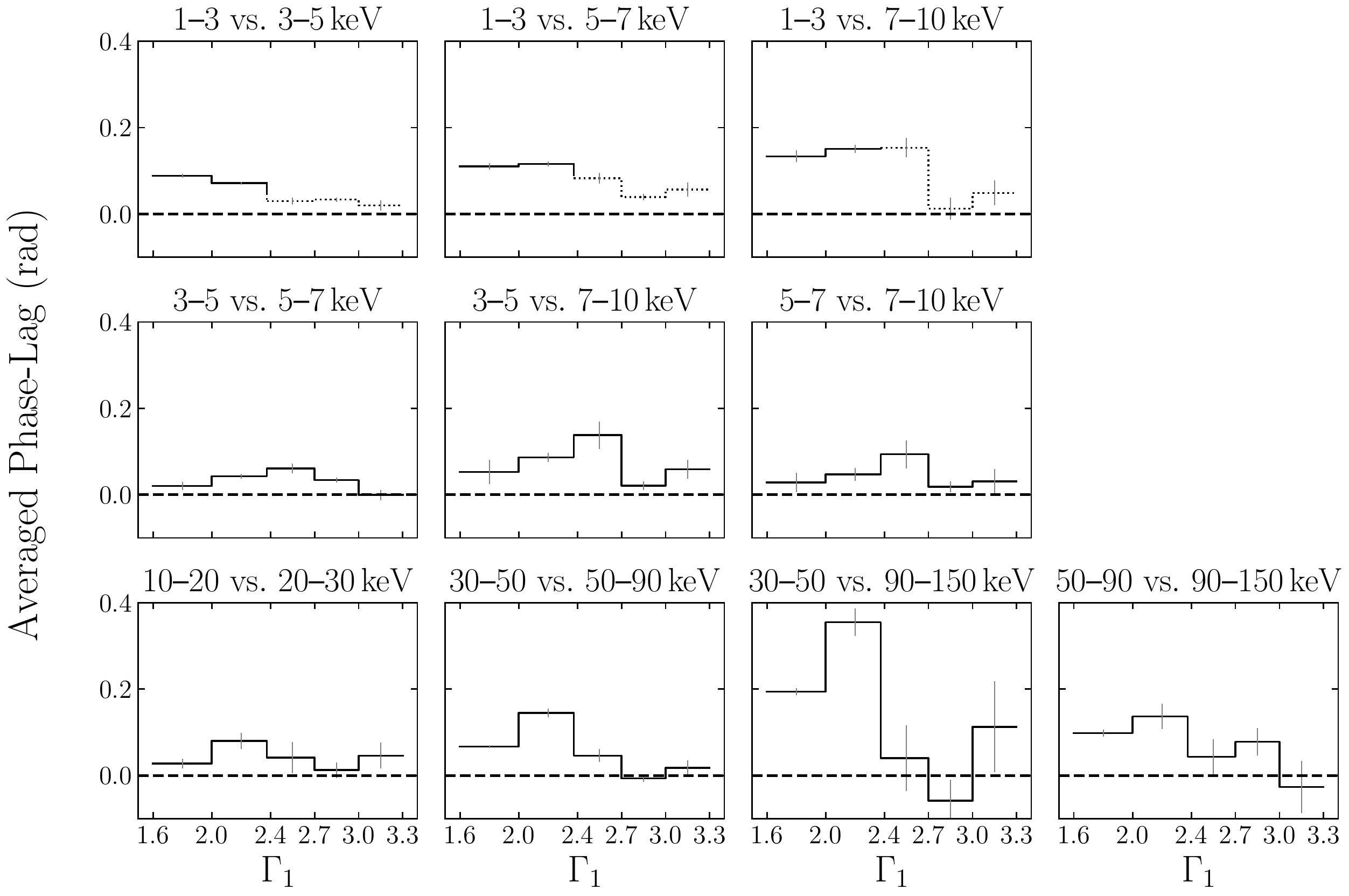}
    \caption{Similar to Fig.~\ref{f-coh}, but with the evolution of the averaged phase-lag versus the states (classified by the soft photon index $\Gamma_1$). In the first three panels, the averaged phase-lags with $\Gamma_1 \geq 2.4$ are denoted by dotted lines, as we know that the loss of coherence is due to the different emission origins. }
    \label{f-phase-lag}
\end{figure*}

On the right-hand side of Fig.~\ref{f-coh-lag-LE}, we show the frequency-dependent phase-lags between different energy bands of LE, but using a different color map. Hard lags are denoted by red, while soft lags are denoted by blue. For ME and HE data, the coherence and the phase-lags are noisier and not suitable to present via this approach. Instead, we show the averaged coherence function and the averaged phase-lags taken over the frequency band and $\Gamma_1$ band. Inspired by the coherence evolution in 1--3 vs. 3--5\,keV band, we divide the soft photon index $\Gamma_1$ into five sub-bands: 1.6--2.0, 2.0--2.4, 2.4--2.7, 2.7--3.0, and 3.0--3.3, consistent with the state definitions in Sect.~\ref{subsect:spec-discussion}. We first rebin the raw coherence and phase-lag spectra to a logarithmically spaced frequency grid with $\mathrm{d}f/f = 0.15$, and average the coherence and the phase-lag in the 0.0625--6\,Hz frequency range in each $\Gamma_1$ band. Coherence greater than 1 (mostly seen in the HE data) shall be corrected as 1 to cancel the effect of strong correlation when averaging. The results are shown in Fig.~\ref{f-coh} and Fig.~\ref{f-phase-lag}. 

\section{Discussion of the timing properties}
\label{sect:discussion_1}

\subsection{The total fractional rms}
\label{subsect:discussion_1-rms}

We show both, corrected and uncorrected rms, in Fig.~\ref{f-index1vsrms} in order to underline the differences between them. For LE, the signal-background ratio is higher than 75\%, and the differences between the two rms estimates can be ignored as clearly seen in the upper row Fig.~\ref{f-index1vsrms}. However, when a non-negligible fraction of photons come from the background, they inevitably smear the power spectrum and then rms. In order to obtain the true trends of total rms using the available data, our assumption can be stated as follows: 1) The raw rms is true if the background has the same intrinsic variability as the source; 2) The corrected rms is true if the background rate is intrinsically invariant and does not contribute to the total variability at all; 3) The true value should lie between the two extreme cases above: the background has its own intrinsic variability, but in all cases the variability from the source dominates. Thus, the true total rms should be between the two trends (denoted by the two orange solid lines) in all panels in Fig.~\ref{f-index1vsrms}. Along with better quality of the data, the two curves that represent the two extreme cases become closer, and the uncertainty range of the true trend is smaller. 

As shown in Fig.~\ref{f-index1vsrms}, the macroscopic positive trends between the total rms and $\Gamma_1$ in hard X-ray bands in the soft state challenge the well-known statement that, due to the domination of the thermal emission in the soft state, the total rms is much lower than that in the hard state~\citep[e.g.,][]{belloni_2005, heil_2015a}. We note that this statement is only true when the soft X-ray band that thermal photons dominate in the soft state is contained (e.g., 1--3\,keV band or 1--10\,keV band). The exact reason will be discussed in Sect.~\ref{subsect:discussion_1-PSDevo}. For other harder bands where Comptonized photons dominate even in the soft state, indeed we see a slightly negative trend when $\Gamma_1 \lesssim 2.7$, and a positive trend when $\Gamma_1 \gtrsim 2.7$. 

The total rms works as an integral of the PSDs over all available frequencies. Different fractional rms values can doubtlessly tell that the properties of the source radiation are different. But even with a flat correlation of rms-$\Gamma_1$, e.g., as shown in the 5.0--7.0\,keV band in Fig.~\ref{f-index1vsrms}, the timing properties in soft/hard states are still quite different. These properties can be revealed by the PSD evolution with spectral shapes, which is presented and discussed in Sect.~\ref{subsect:timing-PSDevo} and Sect.~\ref{subsect:discussion_1-PSDevo}. 

\subsection{The PSD evolution}\label{subsect:discussion_1-PSDevo}

Our work expands on the previous \textsl{RXTE}-based analysis that covered the 2.1--15\,keV range \citep{boeck_2011,grinberg_2014} to both lower and higher energies. Therefore, our maps allow to better trace both the variability of the disk that contributes to the lowest energies, and the variability of the high energy part of the power law. 

In Fig.~\ref{f-PSD-evo}, we see a clear boundary at $\Gamma_1 \simeq 2.7$. Before this boundary, there are at least two pronounced variability components in the 1--10\,Hz range. Their frequency increases when the source softens. 

The central frequencies are approximately independent of the energy bands~\citep[for similar results recently obtained by other BHBs, see e.g.,][]{ma_2021}. The power spectra of \cyg in the hard/intermediate state are comparatively smooth and without narrow features~\citep[see e.g.,][]{axelsson_2005, axelsson_2006, rapisarda_2017}. While we acknowledge that there is some confusion in the literature with the broad variability components having been referred to as QPOs in past analyses~\citep{gilfanov_1999, shaposhnikov_2006}, the behavior of the observed variability components does not agree with the established definition of low frequency QPOs~\citep[see e.g.,][]{ingram_2019}. 

Following \citet{grinberg_2014}, we label the variability component with a lower central frequency as ``Component 1'' and that with a higher central frequency as ``Component 2''. We can see a gradual growth of the strength of Component 2, but a fading Component 1 when the X-ray bands becomes harder. The Component 2 becomes dominant above $\sim$ 10\,keV. Here, we can for the first time trace this behavior above $\sim$ 15\,keV. 

We are able to trace the variability components, especially the Component 2, in the hard state and their increasing frequency with an increasing $\Gamma_1$ to at least the 50--90\,keV range. This is especially clearly visible in Fig.~\ref{f-PSD-evo-he} where we re-scale the map colors to account for the overall low variability detected in the HE instrument. 

\begin{figure*}
    \centering
    \includegraphics[width = 0.99\textwidth]{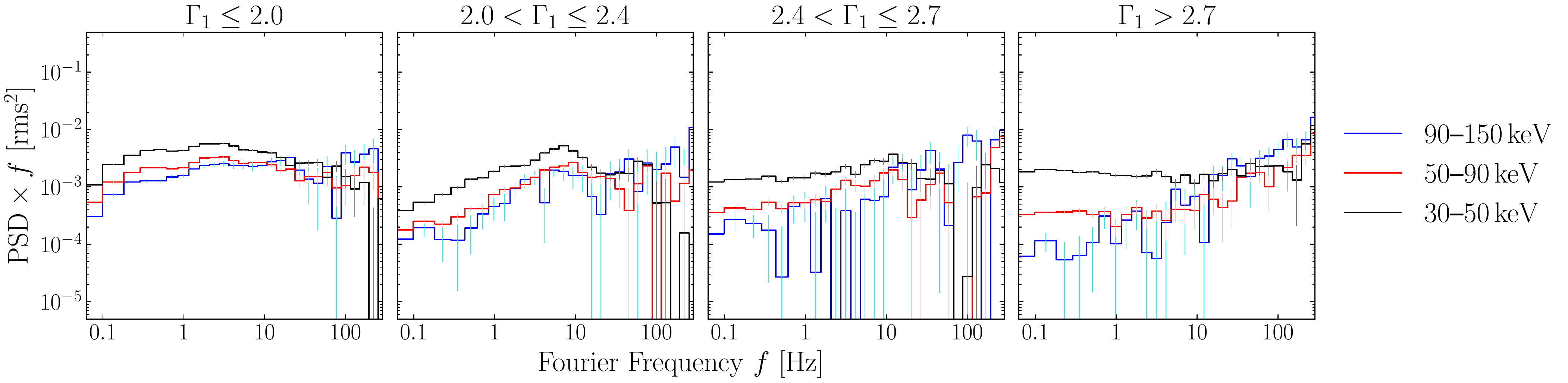}
    \caption{Average power spectra calculated from light curves obtained from the HE instrument in different energy bands in four different $\Gamma_1$ ranges. The PSDs for the 30--50\,keV band, 50--90\,keV band, and 90--150\,keV band are denoted with black/red/blue lines, respectively. The PSDs are rebinned to $\mathrm{d}f/f = 0.3$. }
    \label{f-he-avg-psds}
\end{figure*}

The maps in the 90--150\,keV range do not show a clear pattern. This band is especially interesting since the spectral cutoff in the hard intermediate state is around $\sim$ 150\,keV~\citep[][and our fits in this analysis]{wilms_2006}, with a polarized hard tail present above these energies~\citep{Laurent_2011a, Jourdain_2012a, Rodriguez_2015a, Cangemi_2021a}. To enhance the signal, we calculate the average PSDs for four states defined by $\Gamma_1$ values, the hard state ($\Gamma_1 \leq 2.0$), the hard intermediate state ($2.0 < \Gamma_1 \leq 2.4$), the soft intermediate state ($2.4 < \Gamma_1 \leq 2.7$), and the soft state ($\Gamma_1 > 2.7$) (see Fig.~\ref{f-he-avg-psds}). However, the power spectra in the 90--150\,keV band are still too noisy to allow firm conclusions. We do see an overall decrease in variability, in agreement with previous results using \textsl{INTEGRAL}/SPI~\citep{Cabanac_2011a}.

In the prior studies on \cyg, the variability components seen in the hard and intermediate state can be well modeled by usually two Lorentzians with different central frequencies~\citep{nowak_2000}. Consistent with e.g., \citet{cui_1997_temporal}, \citet{gilfanov_1999}, \citet{pottschmidt_2003}, \citet{axelsson_2006}, \citet{shaposhnikov_2006}, and \citet{boeck_2011} that studied the evolution of the behavior of these components, we find that their central frequencies increase as the source softens. The strength of components in 1--10\,Hz band directly determines the total fractional rms that is shown in Fig.~\ref{f-index1vsrms}. For those bands above 3\,keV, the total rms slightly drops during the hard-to-intermediate transition, where $1.6 \lesssim \Gamma_1 \lesssim 2.7$. But the variability in 1--3\,keV band is additionally influenced by the thermal photons from the accretion disk. The thermal photons that present red-noise spectra in the PSDs make the variability of the smooth components fade more quickly than the harder bands. That is the exact reason why we see a continuously decreasing value of the total rms only in 1--3\,keV band or 1--10\,keV band during the softening of the state. 

When the source continues its transition to the soft state, crossing the boundary of $\Gamma_1 \simeq 2.7$, the power spectrum of hard X-ray bands suddenly changes to a red-noise spectrum with $f^{-1}$ slope below 10\,Hz. A similar behavior of the PSD evolution has also been reported by \citet{grinberg_2014} at energies up to 15\,keV and is traced here for the first time at higher energies, where the power law is the dominating spectral component even in the soft state. The sudden change of the PSD shape at $\Gamma_1 \simeq 2.7$ is the reason why the total rms has a positive tail versus $\Gamma_1$ in the hard X-ray bands. 

In the soft state of \cyg, with $\Gamma_1 \gtrsim 2.7$, the PSD shape shows a red-noise spectrum with $f^{-1}$ slope below 10\,Hz~\citep[see e.g.,][]{cui_1997_rossi,  cui_1997_temporal, gilfanov_2000, churazov_2001, pottschmidt_2003, axelsson_2005, grinberg_2014}. \citet{cui_1997_temporal} and \citet{churazov_2001} also state that the red-noise variability becomes stronger when our focused energy range moves to the harder X-ray bands. This is confirmed by the higher total rms shown by the positive tails of the trends of the corrected rms in Fig.~\ref{f-index1vsrms} and the brightness of the PSD shapes in Fig.~\ref{f-PSD-evo}. 

\subsection{The coherence function and the phase-lags}
\label{subsect:discussion_1-coh-phlag}

Fig.~\ref{f-coh} shows that in the soft state where $\Gamma_1 \geq 2.7$, the HE data ($\geq 30$\,keV) and the LE data involving 1--3\,keV band tend to have worse coherence than those with lower $\Gamma_1$s. The loss of coherence in LE data is interpreted by the domination of the thermal photons from the disk in the 1--3\,keV band. But the loss in HE data is mainly due to the low signal-background ratio when $\Gamma_1 \geq 2.7$. Here we note that \citet{grinberg_2014} do not find similar loss of coherence due to the thermal photons with \textsl{RXTE}/PCA data, as the lower limit of their softest energy band is 2.1\,keV, which in our case is 1.0\,keV. In order to check this supposition, we also extract the event lists from LE for photons of 2.1--4.5\,keV and compute the coherence involving this new energy band. We do not see evident loss of coherence in the soft state. This fact indicates that the thermal photons may dominate the 1--3\,keV band, but cannot dominate the 2.1--4.5\,keV band in the soft state of \cyg. 

Besides, in those bands where the coherence can be well measured, we can see a coherence decrease where $ 2.4 \lesssim \Gamma_1 \lesssim 2.7$. \citet{grinberg_2014} report the similar loss of coherence when $2.35 \lesssim \Gamma_1 \lesssim 2.65$ by analyzing the \textsl{RXTE}/PCA data on \cyg. The origin of the loss of coherence in this soft intermediate state has not been ascertained. Considering that the jet emission in such state becomes weak and unstable or even hardly detectable~\citep[e.g.,][]{fender_2004, boeck_2011, corbel_2013}, the loss of coherence is highly possibly connected with the unstable jet base. 

Since the different origins of the emission in 1--3\,keV band in the hard and the soft state are confirmed by previous spectral and timing analyses, we denote those phase-lags where $\Gamma_1 \geq 2.4$ with dotted lines in Fig.~\ref{f-phase-lag}, and do not discuss them further, because there the quality of the linear coherence between two light curves are bad. 

Fig.~\ref{f-coh-lag-LE} shows that the dominant hard lags contribute from the 1--10\, Hz band when \cyg is in the intermediate state. In the hard state, hard lags tend to be generated at slightly lower frequencies, in comparison with those in the intermediate state. This consistency of the central frequencies of the variability components and the dominant contribution of the frequency-dependent phase-lags indicates that the lags of the hard photons are likely related to the variability components. However, the phase-lags in the soft state in general stochastically fluctuate around 0, resulting in a sudden drop of the averaged phase-lags when $\Gamma_1 > 2.7$, which is shown in Fig.~\ref{f-phase-lag}. 

Most of the averaged phase-lags in Fig.~\ref{f-phase-lag} are positive, which means that in our data, the hard photons usually lag the soft photons. Similar hard lags observed on \cyg are also reported, e.g., by \citet{cui_1997_temporal, nowak_1999, pottschmidt_2003, boeck_2011, grinberg_2014}. 

We are particularly interested in the averaged phase-lags between the adjacent energy bands (e.g., 3--5 vs. 5--7\,keV band or 10--20 vs. 20--30\,keV band), because the phase-lags between the nonadjacent bands can be estimated by summing the phase-lags between the adjacent bands in one $2\pi$ period, if the coherence maintains good. From the 3--5 vs. 5--7\,keV band, 5--7 vs. 7--10\,keV band, and 10--20 vs. 20--30\,keV band in Fig.~\ref{f-phase-lag}, we see that the averaged phase-lags in 0.0625--6\,Hz band always increase as the state softens, where $\Gamma_1$ transitions from 1.6 to 2.7. But when $\Gamma_1$ continues to increase, the averaged phase-lags will decrease. In our work, the maximum lags are observed in the soft intermediate state, where $2.4 \lesssim \Gamma_1 \lesssim 2.7$. \citet{grinberg_2014} reported the lag evolution with \textsl{RXTE}/PCA data on \cyg, with their maximum values at $\Gamma \simeq 2.65$ consistently, but using the form of time-lags, suggesting that the geometry of the corona has a sudden change, which was indicated by the sudden change of the PSD shape as well. 

Intriguingly, this positive correlation between the averaged phase-lags and $\Gamma_1$ no longer holds in the 1--3 vs. 3--5\,keV band, when the source transitions from the hard to hard intermediate state, where $1.6 \lesssim \Gamma_1 \lesssim 2.4$, as shown in Fig.~\ref{f-coh-lag-LE} and Fig.~\ref{f-phase-lag}. Fig.~\ref{f-coh-lag-LE} indicates that the dominant hard lags are contributed from the hard and hard intermediate state, within the frequency band of the variability components. In Fig.~\ref{f-phase-lag} we can see that the phase-lags decrease slightly as $\Gamma_1$ increases. The coherence there is good and the measured phase-lags are accurate. This phenomenon may suggest that in the hard state, although both the 1--3\,keV band and the 3--5\,keV band are both dominated by powerlaw photons, the 1--3\,keV band is more likely to be dominated by the seed photons, which are possibly produced by the thermal black-body radiation of the accretion disk or the synchrotron radiation in the jet base; while the 3--5\,keV band is dominated by the Comptonized photons up-scattered from the seed photons in softer X-ray bands. 

\section{Variability properties of \cyg in a larger context}\label{sect:discussion_2}

So far, we have discussed the individual variability properties of \cyg (Sect.~\ref{subsect:discussion_1-rms}, \ref{subsect:discussion_1-PSDevo}, \ref{subsect:discussion_1-coh-phlag}). We are now putting our results in the context of mechanisms of the state transition, the disk-corona model, and the plausible scenarios of accretion in BHB systems. 

\subsection{Comparison with the behaviors of other BHBs}\label{subsect:discussion_2-comparison}

\cyg is a persistently luminous BHB, fed by the wind of its companion. However, the majority of the BHBs are transient. Although we expect similar accretion scenarios for both kind of systems, it is still meaningful to compare the phenomenology observed for \cyg with other known BHBs. 

The total fractional rms, which is the integral of PSDs over all frequencies, decreases as the state softens~\citep{belloni_2005}. This is also observed in transient and other persistent BHBs~\citep[e.g.,][]{remillard_2006}. However, as discussed in \ref{subsect:discussion_1-rms}, we should always outline the chosen energy band when defining the total rms, as different energy bands  give different trends. 

The evolution of the power spectra during state transitions has been reported in other BHBs, e.g., GX~339$-$4 \citep{zdziarski_2004, motta_2011, shui_2021}, GRS~1915$+$105 \citep{soleri_2008, ueda_2009}, XTE~J1550$-$564 \citep{sobczak_2000, rodriguez_2004, rao_2010}, 4U~1630$-$47 \citep{tomsick_2000}, and Swift~J$1753.5-$0127 \citep{soleri_2013, bu_2019}. Most of the transient BHBs show narrow features in their power spectra, classified into type A, B, and C LFQPOs by their central frequencies, time-lags, and the quality factor \citep{casella_2005, motta_2015}. The central frequencies of the LFQPO components rise when the state of the BHB softens, and drop when returning back to the hard state~\citep{vignarca_2003}. \citet{sobczak_2000} also reported an exception, GRO~J1655$-$40, which shows a negative correlation of the central frequency of the 14--28\,Hz LFQPO versus the powerlaw index, but a positive correlation with the disk flux, during a transition from very high state to the usual soft state. However, \citet{motta_2012} confirmed that the central frequencies of the LFQPO components still increase during the hard-to-soft transition. We note here that the transition from the very high state to the soft may not follow the regulation we talk about, and GRO~J1655$-$40 is also special as the high-frequency quasi-periodic oscillation (HFQPO) and more than two LFQPO components can be observed~\citep{strohmayer_2001, belloni_2012}. 

When those sources complete their transition from the hard state to the soft state, their power spectra with narrow or smooth peaks will be substituted by a red-noise power spectra with $f^{-1}$ slope~\citep[see e.g.,][]{miyamoto_1994, remillard_2006, ueda_2009, klein-wolt_2008, mao_2021}, although \cyg generally shows a higher overall variability than transients in the the soft state \citep[e.g.,][]{heil_2015a}. 

There are fewer papers talking about the higher-order timing properties (the coherence function and phase/time-lag). \citet{reig_2018} report the growths of the hard lag of 8 BHBs when the sources are undergoing a hard-to-intermediate transition. \citet{altamirano_2015} additionally find the sudden decrease of the averaged phase-lag at the transition point between the intermediate state and the soft state in GX~339$-$4, which is consistent with \citet{grinberg_2014} and our results. 

\subsection{Current models to describe timing properties}\label{subsect:discussion_2-timing}

All the timing studies presented and discussed in Sect.~\ref{sect:timing} and \ref{sect:discussion_1} conclude that the variability components are strongly correlated with the mechanism of the transition between the hard and the soft state. Thus, the interpretation of their origin is crucial to reveal the accretion scenario. 

\citet{boeck_2011} reported a fast state transition in \cyg within less than 2.5\,h, but with their available data, they could only reach the soft intermediate state with $\Gamma_1 < 2.7$. The sudden change in the PSD from distinct variability components to red-noise spectrum has not been observed. 

The variability components seen in the PSDs of \cyg are smooth, with a quality factor $Q \lesssim 1$ modeled by two Lorentzian curves~\citep[e.g.,][]{boeck_2011}. Therefore, the models developed for the LFQPOs, e.g., the Lense-Thirring precession~\citep{bardeen_1975} or the accretion-ejection instability~\citep{tagger_1999}, are not suitable in our case. Besides, even for the narrow LFQPOs commonly seen in other transient BHBs, we do not yet have a satisfying physical model to explain the spectral and timing behaviors consistently (see~\citealt{ingram_2019} for a recent review). 

Among those models, a large fraction of them assume a truncated disk/hot inner flow geometry. In this scenario, the hard-to-soft transition happens when the truncation radius between the standard disk and the hot accretion flow decreases, which means an increasing characteristic frequency for all variabilities. Fruitful developments under this scenario have been done so far. For instance, \citet{ingram_2009} consider the precession of a hot flow inside a truncated disk, which is able to explain why the observed maximum frequency is almost constant for all BHBs. \citet{ingram_2011} further include the effect of the local fluctuations in the mass accretion rate into the model, which can reproduce the PSD shapes that commonly seen by real data. Recently, \citet{kawamura_2022} consider propagating mass accretion rate fluctuations to map the \textsl{NICER} data of MAXI~J1820$-$070, which successfully explained the observed two-humps power spectra and the lag-frequency spectrum, supporting that the disk is truncated at a few tens of the gravitational radii. 

While in practice, the inner edge of the accretion disk is usually measured by spectral fitting. However, some obtained results challenge this truncated-disk scenario in the hard state, both via the iron line method~\citep[e.g.,][]{garcia_2015, buisson_2019, sridhar_2020, wang_2021} and the continuum fitting method~\citep[e.g.,][]{miller_2006, reis_2009, reis_2010}, suggesting that even for the hard state, the inner edge of the disk should be close to ISCO. A recent spectral study of \cyg with a physical model set done by \citet{feng_2022}, uses two fixed spin and inclination values suggested by \citet{zhao_2021}, concluding that with both two parameter settings, there is no strong correlation between the state and the inner radius of the disk. If, as indicated by these results, the true mechanism of state transition between the hard and soft state may be independent of the inner edge of the disk, then the evolution of the typical frequencies of variability components in the hard and intermediate states as observed here cannot be explained by a change in the inner disk radius or a precession motion of a flow that changes in size. At this stage, we are inclined to consider the variability component as a geometrical effect originating from the corona, but examining each model quantitatively with our data is beyond the scope of this paper.

In the soft state, the power spectrum is dominated by red noise for all energy bands. The shapes of the PSD are similar, except for the normalization of the $f^{-1}$ powerlaw. A plausible explanation is that the Comptonization process is completely triggered by the thermal photons, indicating that the initial structure of the corona is destroyed. Hence the disappearance of characteristic frequencies of the powerlaw emission, which is presented by variability components in the hard/intermediate states. The broadband noise seen in the power spectrum can be well explained by the fluctuation propagation model~\citep[e.g.,][]{lyubarskii_1997, arevalo_2006}. 

\subsection{The geometry of the corona during the state transitions}\label{subsect:discussion_2-corona}

In order to probe the corona geometry, we first need to empirically explain why the intensity of the reflection component is stronger in the soft state than in the hard state. We partly adopt the interpretation of both \citet{petrucci_2001} and \citet{shaposhnikov_2006}: when the state softens from hard to intermediate, the spatial size of the corona shrinks, and the corona may move closer to the accreting center. This geometrical configuration also matches the apparent phenomenon that the disk component gradually dominates the soft X-ray bands during the process of state softening. Depending on a different technique, the reverberation lag model, \citet{kara_2019} provide similar conclusions~\citep[but also see e.g.,][]{wang_2021}. 

In the hard state, the corona with a higher optical depth can cover the central part of the accretion system, absorbing most of the seed photons and Comptonizing them to produce a powerlaw emission. According to \citet{petrucci_2001}, the reflection component is highly probable to be Comptonized by the large-sized corona as well. Therefore, the reflection feature will be diluted and the measured intensity of the reflection will decrease. When the state softens to the intermediate state, the corona becomes smaller, then part of the seed photons can escape to infinity without being Comptonized. In the soft state with a red-noise-like PSD shape, the corona may shrink to a relatively small size, letting most of the thermal photons escape to infinity, while it only captures and Comptonizes few seed photons to a powerlaw shape. 

The sudden change of the PSD shape and the sudden drop of the phase-lag may even advocate that the shrinking corona is split into pieces at this stage, then losing its intrinsic frequency~\citep{esin_1997, done_2007}. Instead, the emission from these pieces are supported by the seed photons, as their PSD shapes become consistent. This scenario could be indirectly supported by rapid fluctuations of equivalent width of the iron K$\alpha$ line, or the difference between two photon indices $\Delta \Gamma$, when $\Gamma_1 \gtrsim 2.7$, which are compatible with our results as shown in Fig.~\ref{f-hdvsmodelflux} and Fig.~\ref{f-index1vs2}, although the large uncertainties on the $\Delta \Gamma$ do not allow us to firmly conclude that such fluctuations exist. A small-sized corona in the soft state, located at slightly different regions around the accreting center, can possibly produce large differences of the reflection strength. But for a larger corona with higher covering fraction in the hard/intermediate state, such evolution is much smoother. 

\section{Summary \& Outlook}\label{sect:sum}

In this paper, we use a simple phenomenological model to perform the broadband spectral analysis on \cyg with 145 available exposures provided by the HXMT mission. With the best fits obtained, we then performed an energy-resolved timing analysis in the range between 1 and 150\,keV, covering a larger energy range than most previous analyses. We list the most interesting results below: 

\begin{enumerate}
    \item The spectral analysis confirms that the reflection strength becomes higher when the state transitions from hard to soft. 
    \item We offer an explanation for the trends between the total rms versus the soft photon index $\Gamma_1$. The trends depend on the spectral shape and the studied energy band, especially on whether the disk emission dominates the spectrum. 
    \item During the hard-to-soft transition, we see a clear evolution of the central frequencies of the variability components in 1--10\,Hz range with $\Gamma_1$, in particular we are able to trace this behavior up to at least $\sim$ 90\,keV. When the source has fully transitioned to the soft state, the variability component is substituted by the red noise in the power spectrum. 
    \item The high energy photons from the corona are likely to be associated with the variability component with a higher central frequency, suggesting that they are produced from a region very close to the black hole. 
    \item The intrinsic coherence can be used to distinguish the light curves from different states, as the dominant photons are produced via different mechanisms (thermal emission or Comptonization). 
    \item The averaged phase-lag increases during the hard-to-soft transition, but it quickly drops when the transition finishes.
\end{enumerate}

With quasi-simultaneous radio and X-ray monitoring, we may further answer the question of disk-jet coupling, i.e., whether the sudden change of the PSD shape and the recovery of the coherence when $\Gamma \simeq 2.7$ is due to the switching off of the jet emission.

\begin{acknowledgements}
M.~Z. would like to thank the support from China Scholarship Council (CSC 202006100027). This research has made use of NASA’s Astrophysics Data System Bibliographic Services. This research also made use of \texttt{ISIS} functions (\texttt{isisscripts}\footnote{\url{http://www.sternwarte.uni-erlangen.de/isis/}}) provided by ECAP/Remeis observatory and MIT. This work made use of data from the Insight-HXMT mission, a project funded by China National Space Administration (CNSA) and the Chinese Academy of Sciences (CAS). J.~R. acknowledges partial funding from the French space agency (CNES), and the French programme national des hautes énergies. Z.~S. is supported by the National Key R\&D Program of China (2021YFA0718500), the National Natural Science Foundation of China under grants U1838201, U1838202. The material is based upon work supported by NASA under award number 80GSFC21M0002 (CRESST II). 
\end{acknowledgements}

\bibliographystyle{aa}
\bibliography{aa_abbrv, mnemonic, references}

\appendix
\section{Overview of the average PSDs at different spectral shapes}\label{sect:appendix}

\begin{figure*}
    \centering
    \includegraphics[width = 0.99\textwidth]{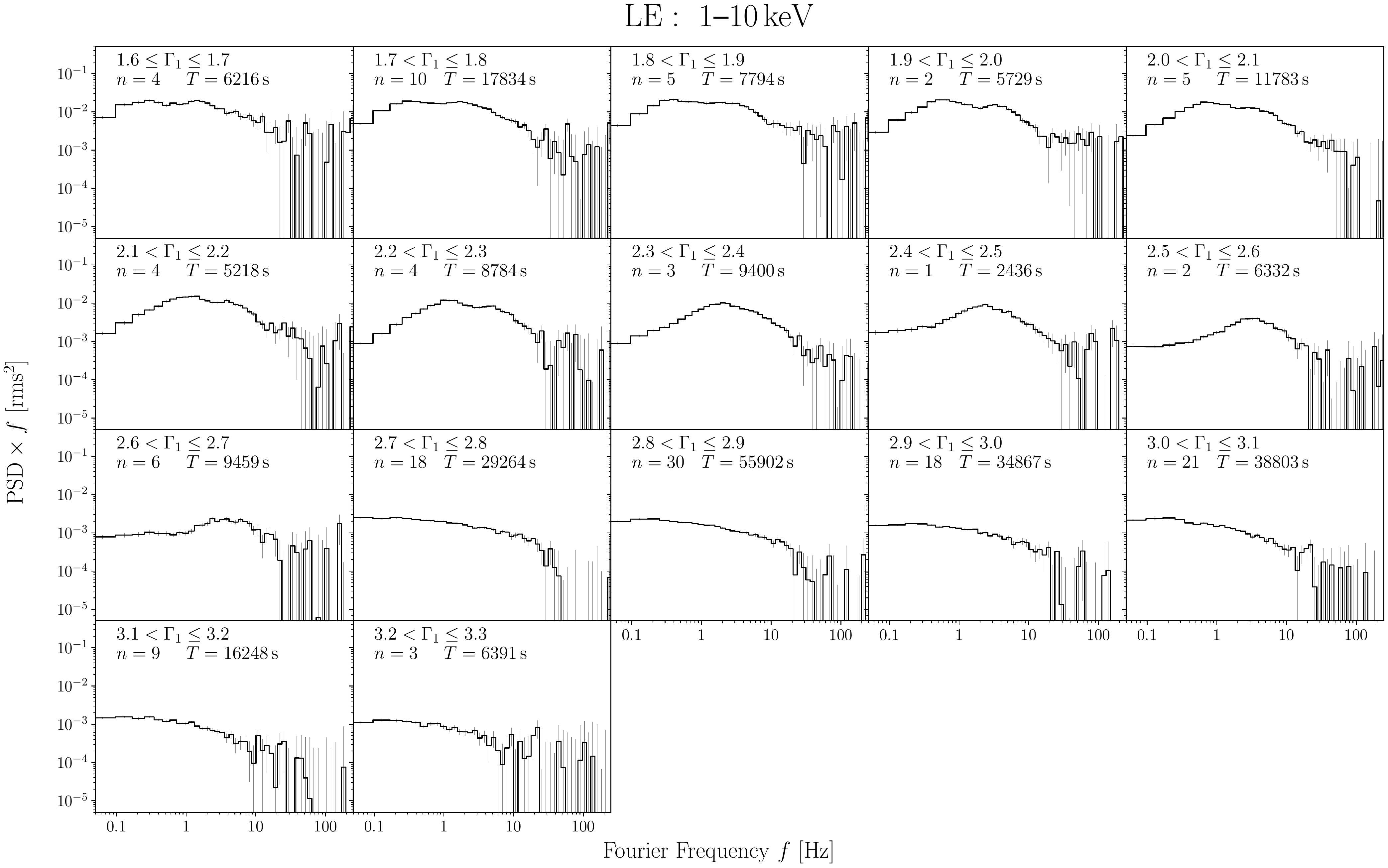}
    \caption{The averaged PSDs of 1--10\,keV band (LE) at different spectral states. Each PSD is the average of all $n$ PSDs falling within the given $\Gamma_1$ interval. We also denote the total exposure time $T$ in GTIs for each $\Gamma_1$ interval. }
    \label{f-psd-le}
\end{figure*}

\begin{figure*}
    \centering
    \includegraphics[width = 0.99\textwidth]{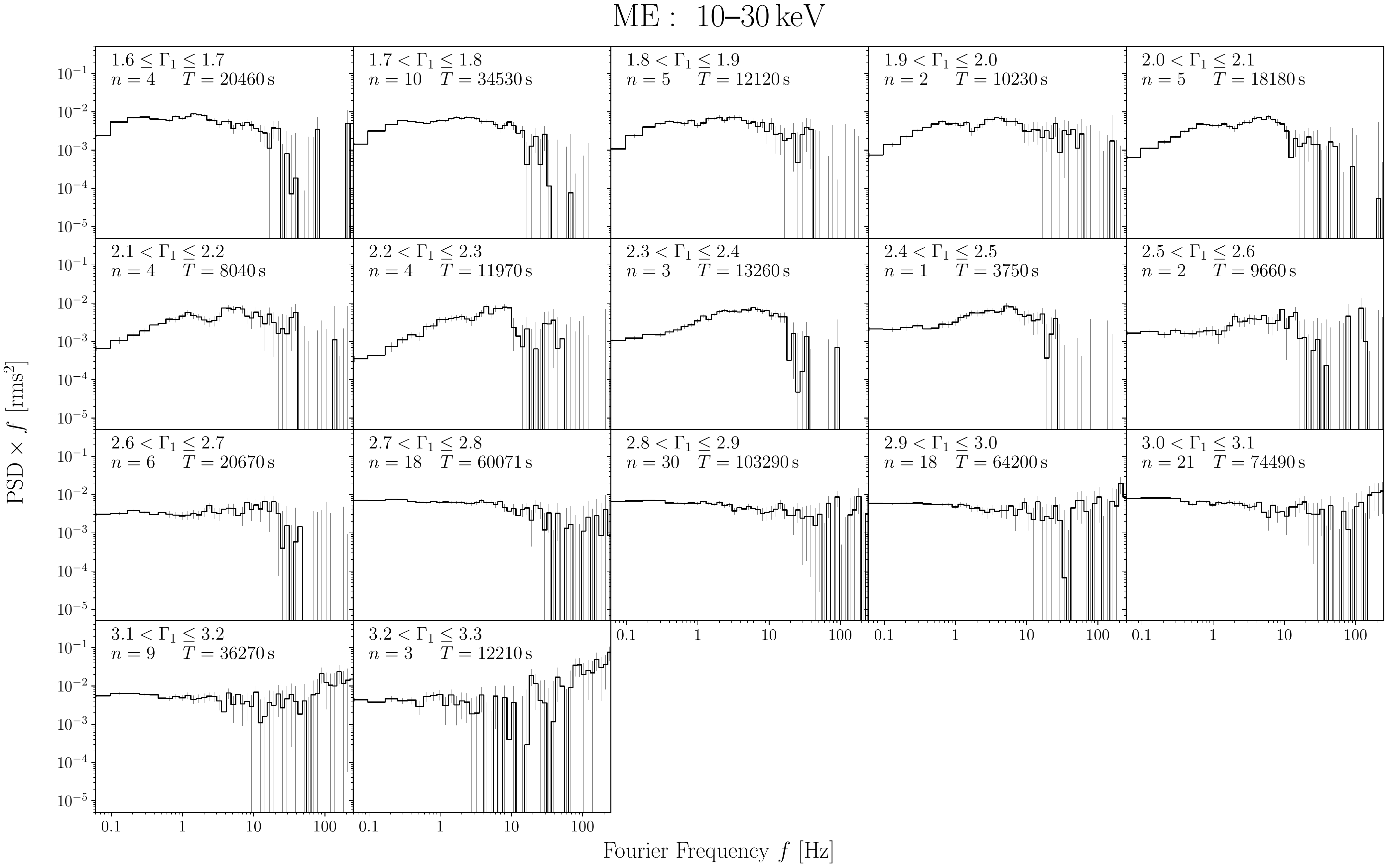}
    \caption{Similar to Fig.~\ref{f-psd-le}, but for 10--30\,keV band (ME). }
    \label{f-psd-me}
\end{figure*}

\begin{figure*}
    \centering
    \includegraphics[width = 0.99\textwidth]{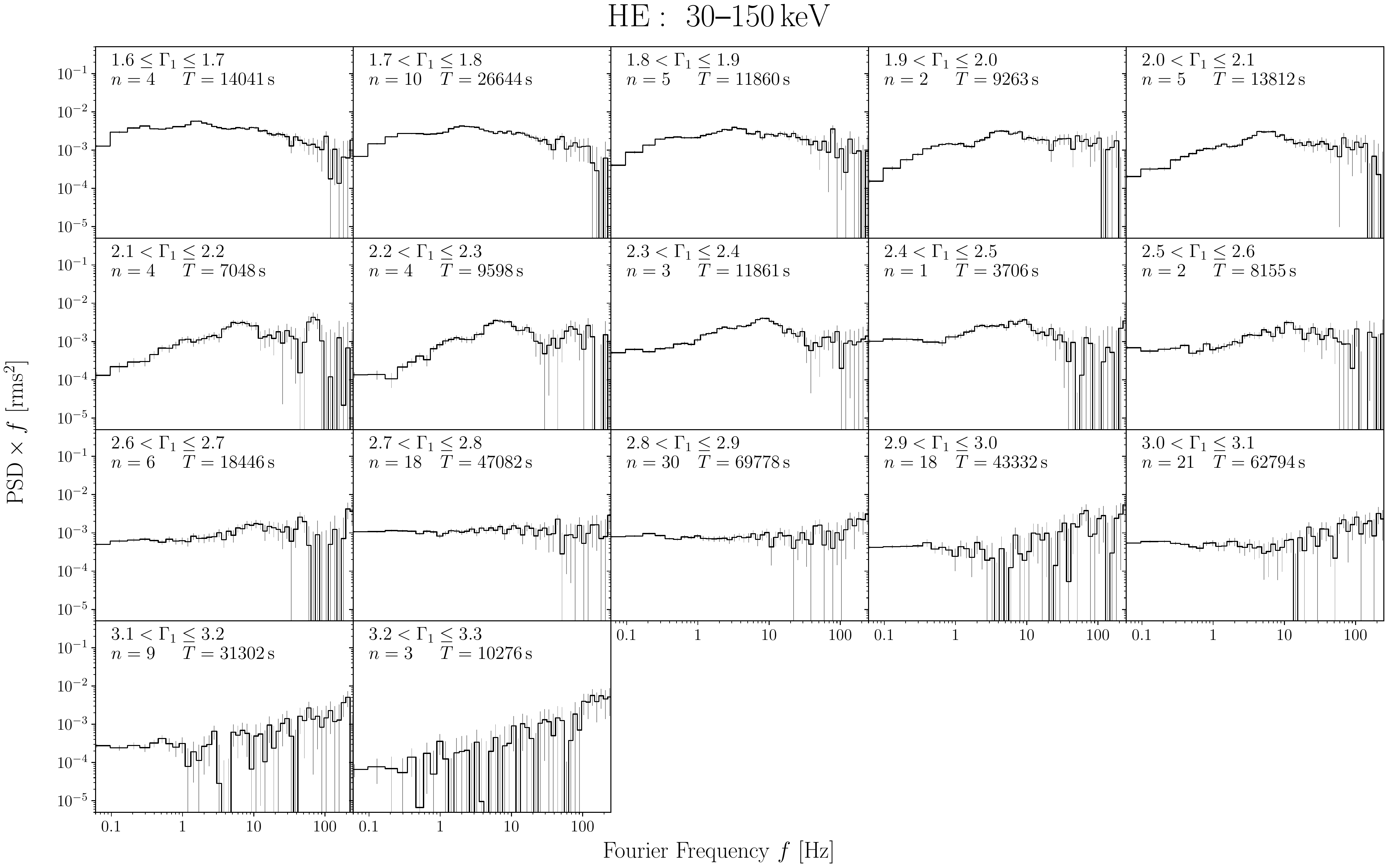}
    \caption{Similar to Fig.~\ref{f-psd-le} and~\ref{f-psd-me}, but for 30--150\,keV band (HE). }
    \label{f-psd-he}
\end{figure*}

\end{document}